\documentclass[journal]{IEEEtran}
\usepackage{epsfig,graphics,subfigure,psfrag,amsmath,cases}
\usepackage{latexsym,amssymb,amsmath,epsfig,subfigure,algorithm}
\usepackage{algorithmic}
\usepackage{color}
\usepackage{url}
\usepackage{scrtime}
\usepackage{datetime}

\author{\authorblockN{ Derrick Wing Kwan Ng,~\IEEEmembership{Member,~IEEE,} Ernest S. Lo,~\IEEEmembership{Member,~IEEE,} \\
and Robert Schober,~\IEEEmembership{Fellow,~IEEE}}\thanks{Manuscript
received November 7, 2013; revised February 11, 2014; accepted March 26, 2014. The review of this paper was coordinated by Prof. C. Yang. This paper has been presented  in part at  IEEE  Globecom 2013 \cite{CN:Kwan_globecom2013}  and  IEEE PIMRC 2013 \cite{CN:Kwan_PIMRC2013}, respectively.
Derrick Wing Kwan Ng and Robert Schober are with the Institute for Digital Communications (IDC),
Friedrich-Alexander-University Erlangen-N\"urnberg (FAU), Germany (email:\{kwan, schober\}@lnt.de).
Ernest S. Lo is with the  Centre Tecnol\`{o}gic de Telecomunicacions de Catalunya - Hong Kong (CTTC-HK) (email: ernest.lo@cttc.hk).  This work was supported in part by the AvH Professorship Program of the Alexander von Humboldt Foundation.}\\
}

\title{Robust Beamforming for Secure Communication in Systems with Wireless Information and Power Transfer}

\date{\ampmtime,\,\today}

\newtheorem{Thm}{Theorem}
\newtheorem{Lem}{Lemma}

\newtheorem{proposition}{Proposition}

\DeclareMathOperator{\Tr}{\mathrm{Tr}}
\DeclareMathOperator{\Rank}{\mathrm{Rank}}

\DeclareMathOperator{\mino}{\mathrm{minimize}}

\newtheorem{Remark}{Remark}

 \newcommand{\qed}{\hfill \ensuremath{\blacksquare}}
\newcommand{\abs}[1]{\lvert#1\rvert}
\newcommand{\norm}[1]{\lVert#1\rVert}

\makeatletter

\newcommand{\Rmnum}[1]{\expandafter\@slowromancap\romannumeral #1@}
\makeatother
\begin{document}
\maketitle\thispagestyle{empty}

\begin{abstract}
This paper considers a multiuser multiple-input single-output (MISO) downlink system with
simultaneous wireless information and power transfer. In particular, we focus on secure communication in the presence of
passive eavesdroppers and  potential eavesdroppers (idle legitimate receivers). We study the design of a resource allocation algorithm minimizing the total transmit power for the case when
the legitimate receivers  are able to harvest energy from  radio frequency signals.  Our design advocates the dual use of both artificial noise and energy signals in providing secure communication and facilitating efficient wireless energy transfer. The algorithm design is formulated as a non-convex optimization problem. The problem formulation takes  into account artificial
noise and energy signal   generation for protecting the transmitted information against both considered types of eavesdroppers when imperfect channel state information (CSI) of the potential eavesdroppers and no CSI of the passive eavesdroppers are available  at the transmitter. Besides, the problem formulation also takes into account different quality of service (QoS) requirements:   a minimum required signal-to-interference-plus-noise ratio (SINR) at the desired receiver; maximum tolerable SINRs at the potential eavesdroppers; a minimum required outage probability at the passive eavesdroppers; and minimum required heterogeneous amounts of power  transferred to the idle legitimate receivers. In light of the intractability of the problem,  we reformulate the considered problem   by  replacing a non-convex probabilistic constraint with a convex deterministic constraint. Then,  a semi-definite programming (SDP) relaxation  approach is adopted to obtain the optimal solution for the reformulated problem. Furthermore,   we propose a suboptimal resource allocation scheme with low computational complexity  for providing communication secrecy and facilitating efficient energy transfer. Simulation results demonstrate the close-to-optimal performance of the proposed  schemes and significant transmit power savings by  optimization of the artificial noise and energy signal generation.

\end{abstract}

\section{Introduction}
\label{sect1}
 \IEEEPARstart{T}{he}
 exponential growth in the demand for high data  rates in wireless communication networks  has
led to a tremendous need for energy. The rapidly escalated energy consumption  not only increases the  operating cost of communication systems, but also raises serious environmental
concerns.  As a result,  green radio communications has received much attention
 in both academia  and  industry \cite{CN:Kwan_globecom2013}--\nocite{CN:WIPT_fundamental,CN:Shannon_meets_tesla,CN:WIP_receiver,JR:MIMO_WIPT,
 JR:WIPT_fullpaper,CN:WIP_energy_beam}\cite{JR:Mag_green}. In particular,   multiple antennas and multicarrier technologies have been proposed  in the literature
 for facilitating energy savings in wireless communication systems.  Unfortunately, the  computational burden and power consumption incurred by the signal processing required for these technologies may be too high for small portable mobile units. As an alternative, multiuser multiple-input multiple-output (MIMO)  has been proposed   where a
transmitter equipped with multiple  antennas services multiple
single-antenna users. However, mobile communication devices are often powered by capacity limited batteries. Hence, the lifetime of mobile devices   remains the bottleneck in the development of ubiquitous wireless communication services.

 The integration of energy harvesting with communication devices is considered  a  promising
alternative  in providing  self-sustainability  to energy limited communication systems \cite{CN:Kwan_globecom2013}--\cite{CN:WIP_energy_beam}.  Hydroelectric, piezoelectric, solar, and wind are the major conventional energy sources for energy harvesting. Nevertheless, the availability of
these natural energy sources is usually  limited by location or
 climate and may be problematic in indoor environments. On
the other hand,    ambient background electromagnetic radiation in radio frequency (RF) is also a viable source of energy for energy scavenging.  More importantly, wireless energy harvesting technology facilitates the possibility of simultaneous wireless information and power transfer. Yet, this new technology introduces a paradigm shift in system and resource allocation algorithm design due to the imposed new challenges.  In \cite{CN:WIPT_fundamental}, the fundamental tradeoff between harvested energy and  data rate was studied. In \cite{CN:Shannon_meets_tesla}, a power allocation algorithm was proposed for near-field communication systems.
 However, the results from \cite{CN:WIPT_fundamental} and \cite{CN:Shannon_meets_tesla} were obtained by assuming that the receiver is able to recycle energy and  decode information from the same received signal which is not possible yet in practice.  Thus,   \emph{power splitting} hybrid receivers and \emph{separated} receivers  were proposed in \cite{CN:WIP_receiver} and \cite{JR:MIMO_WIPT}, respectively,  as a compromise solution. In particular, the \emph{power splitting} receiver splits the received power into two power streams in order to facilitate simultaneous  energy harvesting and information decoding at the receiver.
 In \cite{JR:WIPT_fullpaper}, the authors proposed different power allocation algorithms and showed that introducing power splitting receivers  can improve the energy efficiency of a communication system.    Transmitting an energy signal along with the information signal was proposed   in   \cite{CN:WIP_energy_beam} for
    expediting energy harvesting at the receivers. In practice, the transmitter can increase the energy of the
information carrying signal to facilitate energy harvesting at the receivers.  However, this may also increase the
susceptibility to eavesdropping due to a higher potential for information leakage. Therefore, a new quality of service (QoS) concern on communication security arises in  systems with simultaneous information and power transfer.

In fact,  security is a fundamental problem in wireless communication systems due to the broadcast nature of the wireless medium. Traditionally, communication security relies on  cryptographic encryption employed at the application layer.  However, the commonly used encryption algorithms are based on the  assumption of
 limited  computational capability  at the eavesdroppers  \cite{book:PHY_red,book:PHY_purple}.  Besides, these algorithms assume a perfect secret key management and distribution which may not be possible in wireless networks.  Hence, recently  a large amount of work has been
devoted to information-theoretic physical (PHY) layer
security \cite{Report:Wire_tap}--\nocite{JR:Artifical_Noise1,JR:Kwan_physical_layer,JR:ken_ma_PHY_beamforming}\cite{CN:safe_convex}, as an alternative or complement to cryptographic encryption. The principle of PHY layer security is to exploit the physical characteristics  of the wireless  fading channel for providing perfect secrecy of communication. In his pioneering work on PHY layer security \cite{Report:Wire_tap},  Wyner showed
 that in a wire-tap channel,  a source and a destination can exchange perfectly secure
  information at a strictly positive data rate whenever the source-destination channel enjoys better conditions compared to the source-eavesdropper channel. Hence, exploiting multiple antennas has been proposed in the literature to ensure secure communication. In particular, by utilizing the extra degrees of freedom
offered by multiple antennas,  artificial noise (/interference) is
injected into the communication channel to impair the received signals at the eavesdroppers.
  In \cite{JR:Artifical_Noise1} and
 \cite{JR:Kwan_physical_layer}, power allocation algorithms  were
proposed  for maximizing the ergodic secrecy capacity and   outage secrecy capacity via artificial noise generation, respectively. In \cite{JR:ken_ma_PHY_beamforming}, joint optimization of transmit beamforming and artificial noise  was performed to ensure secure communication in a system with multiple single antenna eavesdroppers. In \cite{CN:safe_convex}, outage-based optimization was proposed to maximize the system secrecy capacity.   However, the energy sources of the  receivers in  \cite{JR:Artifical_Noise1}--\cite{CN:safe_convex} were assumed to be perpetual which may not be possible for energy constrained systems.
  Furthermore, artificial noise generation may consume  a significant portion of the transmit power  to guarantee secure communication   \cite{JR:Artifical_Noise1}--\cite{CN:safe_convex}. On the other hand, the transmitted artificial noise can be harvested by the receivers and the recycled energy can be used to extend the lifetime of power-constrained portable devices.    Yet, the proposed algorithms in  \cite{JR:Artifical_Noise1}--\cite{CN:safe_convex}  do not exploit the dual use of artificial noise for securing the communication and energy harvesting.

The notion  of  secure communication in energy harvesting systems has recently been pursued in
different contexts.
 In  \cite{CN:Kwan_globecom2013}, different power allocation algorithms with  artificial noise generation were designed for power splitting receivers to realize efficient  concurrent power transfer and secure communication.   In \cite{CN:Kwan_PIMRC2013},  a multi-objective approach was used to jointly maximize the  energy harvesting efficiency and minimize the total transmit power while ensuring communication secrecy.  Exploiting an energy signal  to provide secure communication in systems with separated information and energy harvesting receivers was studied in \cite{JR:rui_zhang,CN:rui_zhang_secrecy}. Yet, the assumption of having perfect channel state information (CSI)  of the energy harvesting receivers in \cite{CN:Kwan_globecom2013,CN:Kwan_PIMRC2013,JR:rui_zhang}, and \cite{CN:rui_zhang_secrecy} may be overly optimistic. In fact,  the energy harvesting receivers are not continuously interacting with the transmitter and the corresponding CSI at the transmitter may be outdated even if the channel is only slowly time varying. Besides,  the energy receivers can be  malicious and take advantage  of the system by reporting false CSI (in frequency division duplex (FDD) systems).
In \cite{JR:WIP}, robust beamforming for the case of imperfect CSI was studied for energy harvesting communication systems with one energy harvesting receiver and one information receiver without considering communication security. However, it is unclear if the solution in \cite{JR:WIP} can be extended to multiple energy harvesting receivers while ensuring secure communication. Furthermore,   different types of eavesdroppers may exist  in the system.  For instance,  passive eavesdroppers are illegitimate users who are not registered. They are usually silent to hide their existence from the transmitter. On the other hand, although legitimate users are registered in the system, they may misbehave and eavesdrop the  information signal of other legitimate receivers when they are idle. Thus, idle legitimate receivers are potential eavesdroppers. Yet, the frameworks proposed  in \cite{CN:Kwan_globecom2013,CN:WIPT_fundamental}, \cite{JR:Artifical_Noise1}--\cite{JR:WIP}  do not take into account the coexistence of passive eavesdroppers and potential eavesdroppers.  In addition, these works  do not fully utilize both  artificial noise and  energy signals for providing secure communication and improving wireless energy transfer in the system. Thus, the algorithms proposed in \cite{CN:Kwan_globecom2013,CN:WIPT_fundamental}, \cite{JR:Artifical_Noise1}--\cite{JR:WIP}  may not be able to support the emerging  QoS requirements on secure communication in systems with energy harvesting receivers, especially when  passive eavesdroppers and potential eavesdroppers coexist in the environment.


In this paper, we address the above issues. To this end, the resource allocation algorithm design for secure communication in multiuser multiple-input single-output (MISO) systems with concurrent wireless information and power transfer is formulated
 as a non-convex optimization problem.  The problem formulation encourages the dual use of both artificial noise and energy signals  for facilitating efficient energy transfer and ensuring  communication security.  In particular, both artificial noise and energy signals are used to degrade the interception capabilities  of the eavesdroppers when imperfect CSI of the potential eavesdroppers and no CSI of the passive eavesdroppers are available  at the transmitter.  For obtaining a tractable solution, we reformulate the considered problem   by replacing a non-convex probabilistic constraint with a  tractable convex deterministic constraint.  The
resulting reformulated non-convex optimization problem is solved by a semi-definite programming (SDP) based resource allocation algorithm. Furthermore, we propose  a suboptimal resource allocation scheme  which provides an excellent system performance at  low computational complexity.

 \begin{figure*}
 \centering
\includegraphics[width=5.5in]{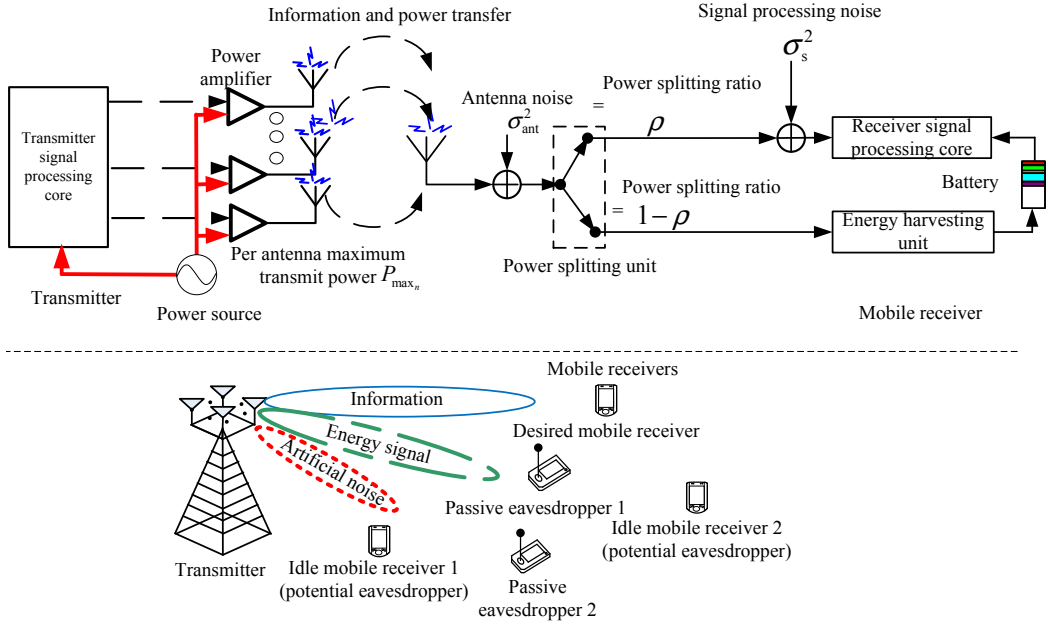}
 \caption{Multiuser system model for $K=3$ legitimate mobile receivers and $J=2$ passive eavesdroppers.  The upper half of the figure illustrates the block diagram of the transceiver
model for wireless information and power transfer.} \label{fig:system_model}
\end{figure*}

\section{System Model}
\label{sect:OFDMA_AF_network_model}
In this section, after introducing the notation used in this paper, we present the adopted multiuser downlink channel model for secure communication with simultaneous wireless information and power transfer.

\subsection{Notation}
  For a square-matrix $\mathbf{X}$,
$\Tr(\mathbf{X})$ denotes the  trace of the matrix
 and $\mathbf{X}\succeq \mathbf{0}$ indicates that
$\mathbf{X}$ is a positive semi-definite matrix. $(\mathbf{X})^H$
and $\Rank(\mathbf{X})$ denote the conjugate transpose and the
rank of matrix $\mathbf{X}$, respectively. $\Big[\mathbf{X}\Big]_{a,b}$ extracts the $(a,b)$-th element of matrix $\mathbf{X}$. Matrix $\mathbf{I}_{N}$
denotes an $N\times N$ identity matrix.  $\mathbb{C}^{N\times M}$ and $\mathbb{R}^{N\times M}$ denote the spaces of $N\times M$ matrices with complex and real entries, respectively.   $\norm{\cdot}$ and $\abs{\cdot}$ denote the
Euclidean norm of a matrix/vector and the absolute value of a complex scalar, respectively.   $\mathrm{Re}(\cdot)$ extracts the real part of a complex-valued input.
$\mathbb{H}^N$ represents the set of all $N$-by-$N$ complex Hermitian matrices. $\lambda_{\max}\big(\mathbf{X}\big)$ and $\lambda_{j}\big(\mathbf{X}\big)$ denote the maximum eigenvalue and the $j$-th eigenvalue  of  Hermitian matrix $\mathbf{X}$, respectively.
The distribution of a circularly symmetric complex Gaussian (CSCG)
vector with mean vector $\mathbf{x}$ and covariance matrix
$\mathbf{\Sigma}$  is denoted by ${\cal
CN}(\mathbf{x},\mathbf{\Sigma})$, and $\sim$ means ``distributed
as". For a real valued  continuous function $f(\cdot)$,
 $\nabla_{\mathbf{X}} f(\mathbf{X})$ represents the gradient of $f(\cdot)$ with respect to matrix $\mathbf{X}$.  ${\cal E}\{\cdot\}$ represents  statistical expectation.

\subsection{Downlink Channel Model}
We consider the downlink of a communication system which consists of a transmitter, $K$ legitimate receivers, and $J$ passive eavesdroppers. The transmitter is equipped with $N_{\mathrm{T}}$ transmit antennas while the legitimate receivers are single antenna devices and are able to decode information and harvest energy from radio signals, cf.  Figure \ref{fig:system_model}. Besides, the $J$ passive eavesdroppers are also equipped with a single antenna, respectively, and their goal is to decode the information transmitted by the transmitter without causing any interference to the communication channel.  There are two types of legitimate receivers in the system; one  desired receiver and $K-1$ idle receivers.
 In each scheduling slot, the transmitter conveys information to the desired receiver and transfers energy\footnote{In this paper,  a normalized energy unit, i.e., Joule-per-second, is adopted.  Therefore,
the terms power and energy are used interchangeably throughout this paper.} to all legitimate receivers simultaneously.
  The $K-1$ idle receivers are supposed to harvest energy from the RF when they are inactive. However, the signals intended for the desired receiver  may be overheard by the idle receivers since all the legitimate receivers are in the range of service coverage.  If the idle receivers are malicious, they may eavesdrop the information signal of the desired receiver. Hence, the idle receivers are potential eavesdroppers which should be taken into account for providing secure communication.  We focus on frequency flat communication channels. The downlink
received signals at the desired receiver, idle receiver (potential eavesdropper) $k$, and  passive eavesdropper\footnote{In practice, the number of passive eavesdroppers is not known at the transmitter. Nevertheless, by assuming a particular value for $J$, we can ensure that the transmitter is able to handle at most $J$ passive eavesdroppers.} $j$ are given by, respectively,
\begin{eqnarray}
y&=&\mathbf{h}^{H} \mathbf{x}+z_{\mathrm{a}},\\
y_{\mathrm{Idle},k}&=&\mathbf{g}_{k}^{H} \mathbf{x} +z_{\mathrm{a},k},\,\,  \forall k\in\{1,\ldots,K-1\},\\
y_{\mathrm{ E},j}&=&\mathbf{l}_{j}^{H} \mathbf{x} +z_{\mathrm{e},j},\,\,  \forall j\in\{1,\ldots,J\},
\end{eqnarray}
where $\mathbf{x}\in\mathbb{C}^{ N_{\mathrm{T}} \times 1}$ denotes the transmitted symbol vector.
$\mathbf{h}^{H}\in\mathbb{C}^{1\times N_{\mathrm{T}}}$ is the channel
vector between the transmitter and the desired receiver
  and
$\mathbf{g}_{k}^{H} \in\mathbb{C}^{1\times N_{\mathrm{T}}}$ is the channel
vector between the transmitter and idle receiver (potential eavesdropper) $k\in\{1,\ldots,K-1\}$. $\mathbf{l}_{j}^{H} \in\mathbb{C}^{1\times N_{\mathrm{T}}}$ is the channel
vector between the transmitter and passive eavesdropper $j\in\{1,\ldots,J\}$. We note that
variables, $\mathbf{h}$, $\mathbf{g}_{k}$, and $\mathbf{l}_{j}$
  include the effects of the multipath fading, shadowing, and path loss
of the associated channels.
  $z_{\mathrm{a}}$, $z_{\mathrm{a},k}$, and $z_{\mathrm{e},j}$  are thermal noises resulting from
  the receive antennas at the desired receiver, idle receiver $k$, and passive eavesdropper $j$, respectively. They are modeled as additive white Gaussian noises (AWGN)  with zero mean and variance $\sigma_{\mathrm{ant}}^2$, respectively,  cf. Figure \ref{fig:system_model}.
  \subsection{Hybrid Information and Energy Harvesting Receiver}
\label{sect:receiver}
Different hardware circuitries \cite{CN:WIP_receiver,JR:RF_energy_circuit} are available for harvesting energy from the RF. The associated system models and the corresponding energy
harvesting efficiencies can be significantly different\footnote{Designing hardware circuitries for harvesting energy from the RF is beyond the scope of this paper. Interested readers may refer
to \cite{CN:WIP_receiver,JR:RF_energy_circuit} for further details. }. Besides, the signal used for decoding the modulated
information cannot be reused for harvesting energy due to hardware limitations \cite{CN:WIP_receiver}.  Therefore,  we do not assume a particular type of energy harvesting circuit such that the resource allocation algorithm design is isolated  from the specific hardware implementation details. In this paper, we adopt hybrid receivers \cite{CN:WIP_receiver,JR:WIPT_fullpaper}
which split the received signal into two power streams with power splitting ratios $1-\rho$ and $\rho$, cf. Figure \ref{fig:system_model}, for harvesting energy and decoding the modulated  information, respectively.  Specifically, the power splitting unit is installed in the front-end of the receiver and is assumed to be a perfect passive analog device; it does not introduce any extra power gain, i.e., $0\le \rho\le 1$, or noise. Indeed, the hybrid receiver is a generalization of traditional information receivers and energy harvesting receivers. In particular,  by
imposing power splitting ratios of $\rho=1$ and $\rho=0$,  the hybrid receiver
reduces to a traditional information receiver and an energy harvesting receiver, respectively. Moreover, the \emph{separated receiver}  proposed in the literature  \cite{CN:Kwan_PIMRC2013,JR:WIP} is also a special case of the considered hybrid receivers.  Furthermore, we
assume that  a rechargeable battery is available  at the receivers for storage of the harvested energy. The stored energy is preserved for future use to extend the lifetime of the receivers. We note that if the amount of harvested energy exceeds the maximum battery capacity, the excess harvested energy will be discarded. On the other hand, we assume that the desired information receiver has stored enough energy to decode the received signal.

\subsection{Channel State Information}
 We assume a Time Division
Duplex (TDD) system with slowly time-varying channels. At the
beginning of each scheduling slot, the legitimate receivers send handshaking/beacon signals\footnote{The legitimate receivers can either take turns to send the handshaking/beacon signals or  transmit simultaneously with orthogonal pilot sequences. } to the transmitter to report their status (service requirements). The information embedded in the handshaking signals facilitates the downlink packet transmission. In particular, the downlink CSI of the transmitter-to-legitimate receiver channels is
obtained through measuring the uplink pilots in the handshaking/beacon signals  via channel reciprocity.
Thus, the transmitter-to-legitimate receiver fading gains, $\mathbf{h}$ and $\mathbf{g}_{k},\forall k\in\{1,\ldots,K-1\}$, can be reliably estimated at the transmitter at the beginning of each scheduling slot with negligible estimation errors.  During the transmission, the desired receiver sends acknowledgement (ACK) packets  to inform the transmitter of successful reception of information packets. As a result, the transmitter is able to refine the estimate of $\mathbf{h}$ frequently via the pilot sequences embedded in each ACK packet. Therefore, we can assume perfect CSI for the transmitter-to-desired receiver  link during the whole transmission period. However, the CSI of the idle receivers becomes outdated during the transmission since there is no interaction between the transmitter and the idle receivers.  To capture this
effect, we use  a deterministic model \cite{JR:Robust_error_models1,JR:Robust_error_models2} for modelling the resulting CSI uncertainty. The  CSI of the link between the transmitter
and idle receiver $k$ is given by
\begin{eqnarray}\label{eqn:outdated_CSI}
\mathbf{g}_k&=&\mathbf{\hat g}_k + \Delta\mathbf{g}_k,\,   k\in\{1,\ldots,K-1\}, \mbox{   and}\\
{\Omega }_k&\triangleq& \Big\{\Delta\mathbf{g}_k\in \mathbb{C}^{N_{\mathrm{T}}\times 1}  :\Delta\mathbf{g}_k^H \Delta\mathbf{g}_k \le \varepsilon_k^2\Big\},\label{eqn:outdated_CSI-set}
\end{eqnarray}
where $\mathbf{\hat g}_k\in\mathbb{C}^{N_{\mathrm{T}}\times 1}$ is the channel estimate of idle receiver $k$ available at the transmitter at the beginning of a scheduling slot. $ \Delta\mathbf{g}_k$ represents the unknown channel uncertainty  of idle receiver $k$ due to the slow timing varying nature of the channel during transmission. For notational simplicity, we define a set ${\Omega }_k$ in (\ref{eqn:outdated_CSI-set}) which contains all possible CSI uncertainties of idle receiver $k$. Specifically, the radius $\varepsilon_k>0$ represents the size of the uncertainty region of the estimated CSI of idle receiver  $k$. In practice, the value of $\varepsilon_k^2$ depends on the coherence time of the associated channel and the time between channel estimation and  packet transmission.

 On the other hand, the passive eavesdroppers, which are also present in the system, are usually silent to hide their existence from the transmitter.
As a result,  the CSI of the passive eavesdroppers cannot be
 measured at the transmitter based on handshaking signals. To facilitate the resource allocation algorithm design, we assume that the transmitter knows that the eavesdroppers are equipped with single antennas. In addition, except for the  variance, the distribution of the channels between the transmitter and the  passive eavesdroppers is assumed to be known at the transmitter.

\subsection{Artificial Noise and Energy Signal Generation}
In order to provide
secure communication and to facilitate energy harvesting at the desired receivers, artificial noise signals and energy
signals are generated at the transmitter. In particular, both signals are able to degrade the channels between
the transmitter and the passive eavesdroppers and act as energy source for energy harvesting. The transmitter chooses the transmit signal vector $\mathbf{x}$ as
\begin{eqnarray}
\mathbf{x}=&&\underbrace{\mathbf{w}s}_{\mbox{desired information signal}}+\underbrace{\mathbf{w}_{\mathrm{E}}}_{\mbox{energy signal}}\notag\\
+&&\underbrace{\mathbf{v}}_{\mbox{artificial noise}},
\end{eqnarray}
where $s\in\mathbb{C}$ and $\mathbf{w}\in\mathbb{C}^{N_{\mathrm{T}}\times 1}$  are the information-bearing signal and the corresponding  beamforming vector for the desired receiver, respectively. We assume without loss of generally that  ${\cal E}\{\abs{s}^2\}=1$. $\mathbf{v}\in\mathbb{C}^{N_{\mathrm{T}}\times 1}$ is the artificial noise vector generated by the transmitter to combat both potential and passive eavesdroppers. $\mathbf{v}$ is modeled as a complex Gaussian random vector with
\begin{eqnarray}
\mathbf{v}\sim {\cal CN}(\mathbf{0}, \mathbf{V}),
\end{eqnarray}
where $\mathbf{V}\in \mathbb{H}^{N_{\mathrm{T}}}, \mathbf{V}\succeq \mathbf{0}$, denotes the covariance matrix of the artificial noise.  The artificial noise signal $\mathbf{v}$ is  unknown to both the legitimate receivers and passive eavesdroppers. On the other hand, $\mathbf{w}_{\mathrm{E}}$ is a Gaussian pseudo-random sequence\footnote{For energy transfer, the energy sequence is not required to be generated by a  Gaussian   pseudo-random source. However, in this paper, a Gaussian pseudo-random energy sequence is assumed to be able to also provide secure communication.} which is used to facilitate efficient energy transfer and is known to all legitimate  (both active and idle) receivers. $\mathbf{w}_{\mathrm{E}}$ is modeled as a complex Gaussian pseudo-random vector with
\begin{eqnarray}
\mathbf{w}_{\mathrm{E}}\sim {\cal CN}(\mathbf{0}, \mathbf{W}_{\mathrm{E}}),
\end{eqnarray}
where $\mathbf{W}_{\mathrm{E}}\in \mathbb{H}^{N_{\mathrm{T}}},\mathbf{W}_{\mathrm{E}}\succeq \mathbf{0}$ denotes the covariance matrix of the pseudo-random energy signal.


\begin{Remark} We assume that energy signal $\mathbf{w}_{\mathrm{E}}$ is a pseudo-random signal which is perfectly known at the transmitter and the desired receivers but not at the passive eavesdroppers.  This can be realized by using
 a  short secret key  as seed information for the  pseudo-random sequence generator used for generating
the energy signal sequences, where the transmitter regularly changes the seeds to prevent the sequence from being cracked by the passive eavesdroppers. The seeds information used at the transmitter can be delivered to the desired receivers securely by
exploiting e.g. the reciprocity of the channels between the transmitter and the legitimated receivers \cite{CN:secrect_key}.
\end{Remark}

\begin{Remark}
We note that unlike  other system models adopted in the literature, e.g. \cite{JR:Artifical_Noise1}--\cite{JR:ken_ma_PHY_beamforming}, the artificial noise and energy  signals in the considered system can be exploited by the legitimate receivers. Although the transmitter could  also solely use the energy of the information carrying signal for supplying  energy to the legitimate receivers \cite{CN:WIP_receiver,JR:WIPT_fullpaper}, increasing the transmit power of the information signal for facilitating energy harvesting  increases the susceptibility to  eavesdropping. As a result, in this paper, we advocate the dual use of artificial noise signals and energy signals in providing simultaneous security and efficient energy transfer.
\end{Remark}
\section{Resource Allocation Algorithm Design}\label{sect:forumlation}
In this section, we first define the channel capacity and secrecy capacity as the system performance metrics. Then, we formulate the corresponding resource allocation algorithm design as an optimization problem. For the sake of notational simplicity, we define the following variables:
$\mathbf{H}=\mathbf{h}\mathbf{h}^H$,  $\mathbf{G}_k=\mathbf{g}_k\mathbf{g}_k^H, k\in\{1,\ldots,K-1\}$, and  $\mathbf{L}_j=\mathbf{l}_j\mathbf{l}_j^H, j\in\{1,\ldots,J\}$.

\subsection{System Capacity and Secrecy Capacity}
\label{subsect:Instaneous_Mutual_information}
 The energy signal $\mathbf{w}_{\mathrm{E}}$ is a Gaussian pseudo-random  sequence which is only known at the legitimate  receivers. Hence, interference cancellation  can be performed at the legitimated receivers for improving the  channel capacity for information decoding. As a result,  given perfect CSI at the
receiver,  the channel capacity (bit/s/Hz)  between the transmitter and the desired receiver
can be rewritten as
\begin{eqnarray}\label{eqn:cap_SIC}
C^{\mathrm{IC}}&=&\log_2\Big(1+\Gamma^{\mathrm{IC}}\Big)\,\,\,\,
\mbox{and}\notag\\
\Gamma^{\mathrm{IC}}&=&\frac{\rho\,\mathbf{w}^H\mathbf{H}\mathbf{w}}
{\rho(\sigma_{\mathrm{ant}}^2+\Tr(\mathbf{H}\mathbf{V}))+\sigma_{\mathrm{s}}^2 },
\end{eqnarray}
where $\Gamma^{\mathrm{IC}}$ is the received signal-to-interference-plus-noise ratio (SINR) at the desired receiver when interference cancellation is performed \cite{CN:WIP_energy_beam}, i.e., $\Tr(\mathbf{H} \mathbf{W}_{\mathrm{E}})$ has been removed. $\sigma_{\mathrm{s}}^2 $ is the signal processing noise power at the receiver\footnote{We assume that the signal processing and thermal noise characteristics are identical for all receivers due to  similar hardware architectures.}, cf. Figure \ref{fig:system_model}.
On the other hand, interference cancellation can also be performed at the idle receivers. Therefore,  the channel capacity between the transmitter and idle receiver (potential eavesdropper) $k$  is given  by
\begin{eqnarray}\label{eqn:cap-eavesdropper}
C_{\mathrm{I}_k}^{\mathrm{IC}}&=&\log_2\Big(1+\Gamma_{k}^{\mathrm{IC}}\Big)\,\,\,\,
\mbox{and}\,\,\\ \label{eqn:SINR_up_idle}
\Gamma_{\mathrm{I}_k}^{\mathrm{IC}}&=&\frac{\rho_k\mathbf{w}^H\mathbf{G}_k\mathbf{w}}
{\rho_k(\sigma_{\mathrm{ant}}^2+\Tr(\mathbf{G}_k\mathbf{V}))+\sigma_{\mathrm{s}}^2 }\notag\\
   &\stackrel{(a)}{\le}& \frac{\mathbf{w}^H\mathbf{G}_k\mathbf{w}}{\sigma_{\mathrm{ant}}^2+\Tr(\mathbf{G}_k\mathbf{V})+\sigma_{\mathrm{s}}^2 }   , \end{eqnarray}
where $\rho_k$ and  $\Gamma_{\mathrm{I}_k}^{\mathrm{IC}}$ are the power splitting ratio and the received SINR at idle receiver $k$, respectively. $(a)$ is due to the fact that $\Gamma_{\mathrm{I}_k}^{\mathrm{IC}}$ is a monotonically increasing function of $\rho_k$. The physical meaning of (\ref{eqn:SINR_up_idle}) is that  idle receiver $k$ gives up the opportunity to harvest energy  and devotes all the received power to eavesdropping.

In practice, the transmitter does not know the location of the passive eavesdroppers.  As a result, we design the resource allocation algorithm assuming an  unfavourable scenario for the location of the passive eavesdroppers. In particular, we assume that passive eavesdropper $j$ is close to the transmitter and is located at the reference distance of the path loss model\footnote{In practice, the reference distance can be treated as a security guard zone and we assume it is known that passive eavesdroppers do not exist inside this zone. }.  Thus, under this scenario, the  capacity between the transmitter and passive eavesdropper $j\in\{1,\ldots,J\}$ is given by
\begin{eqnarray}  \label{eqn:eavesdropper-SINR-bound}
C_{\mathrm{PE}_j}&\stackrel{(a)}{=} &\log_2\Big(1+\Gamma_{\mathrm{PE}_j}\Big)\,\,\,\,
\mbox{and}\,\,\\ \label{eqn:SINR_up_passive} \notag
\Gamma_{\mathrm{PE}_j}&=&\frac{\mathbf{w}^H\mathbf{L}_j\mathbf{w}}{
\Tr(\mathbf{L}_j\mathbf{W}_\mathrm{E})+\Tr(\mathbf{L}_j\mathbf{V})+\sigma_{\mathrm{ant}}^2+\sigma_{\mathrm{s}}^2 }\\
&\stackrel{(b)}{\le}& \frac{\mathbf{w}^H\mathbf{{L}}_j^{\mathrm{UP}}\mathbf{w}}{\Tr(\mathbf{{L}}_j^{\mathrm{UP}}\mathbf{W}_\mathrm{E})+
\Tr(\mathbf{{L}}_j^{\mathrm{UP}}\mathbf{V})+\sigma_{\mathrm{ant}}^2+\sigma_{\mathrm{s}}^2 }\\   \label{eqn:SINR_up_passive-up}
&\stackrel{(c)}{=} & \frac{\mathbf{w}^H\mathbf{\tilde{L}}_j\mathbf{w}}{\Tr(\mathbf{\tilde{L}}_j\mathbf{W}_\mathrm{E})+
\Tr(\mathbf{\tilde{L}}_j\mathbf{V})+\tilde\sigma^2_j },
\end{eqnarray}
where $\mathbf{{L}}_j^{\mathrm{UP}}=\mathbf{{l}}_j^{\mathrm{UP}}(\mathbf{{l}}_j^{\mathrm{UP}})^H$, and   $\mathbf{{l}}_j^{\mathrm{UP}}$  is the channel vector between the transmitter and a passive eavesdropper located at the reference distance. We note that the elements of $\mathbf{{l}}_j^{\mathrm{UP}}$ and $\mathbf{{l}}_j$  capture the joint effect of small scale fading and shadowing in the same manner. Yet, the path loss coefficient contained in $\mathbf{{l}}_j^{\mathrm{UP}}$ is calculated at the reference distance which results in $\norm{\mathbf{{l}}_j^{\mathrm{UP}}}\ge\norm{\mathbf{{l}}_j}$.
 {The equality in $(a)$ is due to the assumption that the energy signal $\mathbf{w}_{\mathrm{E}}$ is a Gaussian random signal.}  The inequality in $(b)$ is due to the fact that the SINR in (\ref{eqn:SINR_up_passive}) is a monotonically increasing function with respect to $\mathbf{{L}}_j$ and $\norm{\mathbf{l}_j^{\mathrm{UP}}}\ge\norm{\mathbf{{l}}_j}$. On the other hand,
   $\mathbf{\tilde{L}}_j=\mathbf{\tilde l}_j\mathbf{\tilde l}_j^H=\frac{\mathbf{{L}}_j^{\mathrm{UP}}}{{\cal E}\{\norm{\mathbf{l}_j^{\mathrm{UP}}}^2\}}$ and   $\tilde\sigma^2_j=\frac{\sigma_{\mathrm{ant}}^2+\sigma_{\mathrm{s}}^2}{{\cal E}\{\norm{\mathbf{l}_j^{\mathrm{UP}}}^2\}}$ in $(c)$ denote the normalized channel matrix and noise power, respectively.  Specifically, $(c)$ in (\ref{eqn:SINR_up_passive-up}) is obtained by dividing\footnote{The normalization facilitates the resource allocation algorithm design in the later parts of this paper.  }  both the nominator and denominator of the upper bound SINR by ${\cal E}\{\norm{\mathbf{l}_j^{\mathrm{UP}}}^2\}$.  Besides, since ${\cal E}\{\norm{\mathbf{l}_1^{\mathrm{UP}}}^2\}={\cal E}\{\norm{\mathbf{l}_2^{\mathrm{UP}}}^2\}=\ldots={\cal E}\{\norm{\mathbf{l}_J^{\mathrm{UP}}}^2\}$, we rewrite $\tilde\sigma_{j}^2$ as $\tilde\sigma^2$ in the sequel without loss of generality.  We note that the passive eavesdroppers are unable to perform interference cancellation to remove $\Tr(\mathbf{L}_j\mathbf{W}_\mathrm{E})$ since the energy sequence $\mathbf{w}_{\mathrm{E}}$ is only known at the legitimate receivers.  With a slight abuse of notation, we reuse variables  $C_{\mathrm{I}_k}^{\mathrm{IC}}$  and $C_{\mathrm{PE}_j}$ to
 denote their upper bounds  by replacing the SINRs $\Gamma_{\mathrm{I}_k}^{\mathrm{IC}}$ and  $\Gamma_{\mathrm{PE}_j}$ in (\ref{eqn:cap-eavesdropper}) and  (\ref{eqn:eavesdropper-SINR-bound}) with their upper bounds in (\ref{eqn:SINR_up_idle}) and (\ref{eqn:SINR_up_passive-up}), respectively.


 Therefore, the maximum achievable secrecy capacity between the transmitter
and the desired receiver can be expressed as \cite{JR:Artifical_Noise1}
\begin{eqnarray}\label{eqn:secrecy_cap}
\hspace*{-3mm}&&\hspace*{-5mm}C_{\mathrm{sec}}\notag\\
\hspace*{-3mm}=&&\hspace*{-5mm}\Big[C^{\mathrm{IC}} - \max\Big\{\underset{k\in\{1,\ldots,K-1\}}{\max} C_{\mathrm{I}_k}^{\mathrm{IC}},
\underset{j\in\{1,\ldots,J\}}{\max}  C_{\mathrm{PE}_j}
 \Big\}\Big]^+.
\end{eqnarray}
$C_{\mathrm{sec}}$ quantifies the maximum
achievable data rate at which a transmitter can reliably send
 secret information to the desired receiver such that
the eavesdroppers are unable to decode the received signal even if the eavesdroppers have infinite computational power for decoding the received signals.

\begin{Remark}
Equations (\ref{eqn:cap-eavesdropper})--(\ref{eqn:SINR_up_passive-up}) reveal that the artificial noise signal, $\mathbf{v}$, and the energy signal, $\mathbf{w}_{\mathrm{E}}$, have different effectiveness in providing communication security. In particular, the artificial noise signal  $\mathbf{v}$ is able to degrade the channels of both idle receivers (potential eavesdroppers) and passive eavesdroppers simultaneously, while $\mathbf{w}_{\mathrm{E}}$ can only degrade the channels of the passive eavesdroppers. On the other hand, both $\mathbf{v}$ and $\mathbf{w}_{\mathrm{E}}$ have the same effectiveness in facilitating    energy transfer to the legitimate  receivers.

%

%

 \end{Remark}

\subsection{Optimization Problem Formulation}
\label{sect:cross-Layer_formulation}
The optimal resource allocation policy, $\Big\{{\mathbf w}^*$, ${\rho}^*$,${\mathbf{W}_{\mathrm{E}}^*}$ ,${\mathbf  V}^*\Big\}$, for minimizing the total power radiated  by the transmitter,  can be
obtained by solving
\begin{eqnarray}
\label{eqn:cross-layer}&&\hspace*{5mm} \underset{{\mathbf{W}_{\mathrm{E}}},\mathbf{V}\in \mathbb{H}^{N_{\mathrm{T}}},\mathbf{w}, \rho
}\mino\,\,\, \norm{\mathbf{w}}^2+\Tr(\mathbf{V})+\Tr(\mathbf{W}_{\mathrm{E}})\nonumber\\
\notag \mbox{s.t.} &&\hspace*{-5mm}\mbox{C1: }\notag\frac{\rho\abs{\mathbf{h}^H\mathbf{w}}^2}{\rho(\sigma_{\mathrm{ant}}^2+\Tr(\mathbf{H}
\mathbf{V}))+\sigma_{\mathrm{s}}^2 } \ge \Gamma_{\mathrm{req}}, \\
&&\hspace*{-5mm}\mbox{C2: }\notag\max_{\Delta\mathbf{g}_k\in {\Omega}_k}\frac{\abs{\mathbf{g}_k^H\mathbf{w}}^2}
{\sigma_{\mathrm{ant}}^2+\Tr(\mathbf{G}_k\mathbf{V})+\sigma_{\mathrm{s}}^2 } \le \Gamma_{\mathrm{tol}_k},\forall k,\\
&&\hspace*{-5mm}\mbox{C3: }\notag\Pr\Big(\hspace*{-0.5mm}{\max_{j\in\{1,\ldots,J\}} \hspace*{-0.5mm} \Big\{\frac{\mathbf{w}^H\mathbf{\tilde{L}}_j\mathbf{w}}{\Tr(\mathbf{\tilde{L}}_j\mathbf{W}_\mathrm{E})\hspace*{-0.5mm}+\hspace*{-0.5mm}
\Tr(\mathbf{\tilde{L}}_j\mathbf{V})\hspace*{-0.5mm}+\hspace*{-0.5mm}\tilde\sigma^2_j }\hspace*{-0.5mm}\Big\}}\hspace*{-0.5mm} \le \hspace*{-0.5mm} \Gamma_{\mathrm{tol}}\hspace*{-0.5mm}\Big)\notag\\
&&\hspace*{1mm}\notag\ge \kappa,\\
&&\hspace*{-5mm}\mbox{C4: }\notag(1-\rho)\eta\abs{\mathbf{h}^H\mathbf{w}}^2+(1-\rho)\eta\Big(\Tr(\mathbf{H}\mathbf{V})\notag\\
&&\hspace*{1mm}+\Tr(\mathbf{H}\mathbf{W}_{\mathrm{E}})+\sigma_{\mathrm{ant}}^2\Big)\ge P_{\min}, \notag\\
&&\hspace*{-5mm}\mbox{C5: }\notag \min_{\Delta\mathbf{g}_k\in {\Omega}_k} \eta\abs{\mathbf{g}_k^H\mathbf{w}}^2+\eta\Big(\Tr(\mathbf{G}_k\mathbf{V})\notag\\
&&\hspace*{1mm}\notag+\Tr(\mathbf{G}_k\mathbf{W}_{\mathrm{E}})+\sigma_{\mathrm{ant}}^2\Big)\ge P_{\min_k},\forall k, \\
&&\hspace*{-5mm}\mbox{C6: }\notag \Big[\mathbf{w}\mathbf{w}^H\Big]_{n,n}  +\Big[\mathbf{V}\Big]_{n,n}+\Big[{\mathbf{W}_{\mathrm{E}}}\Big]_{n,n}\notag\\
&&\hspace*{1mm}\notag\le P_{\max_n}, \forall n\in\{1,\ldots,N_{\mathrm{T}}\}, \\
&&\hspace*{-5mm}\mbox{C7:}\,\, 0\le\rho\le 1,\quad\mbox{C8:}\,\, \mathbf{V},\mathbf{W}_{\mathrm{E}}\succeq \mathbf{0}.
\end{eqnarray}
In C1, $\Gamma_{\mathrm{req}}$ denotes the minimum SINR
 of the desired receiver required for information decoding. This constraint guarantees  that the channel capacity between the transmitter and the desired receiver is $C^{\mathrm{IC}}\ge \log_2(1+\Gamma_{\mathrm{req}})$. Constraint C2 is imposed such that for a given CSI uncertainty set $\Omega_k$, the maximum received SINR at idle receiver (potential eavesdropper) $k$ is less than the maximum tolerable received SINR $\Gamma_{\mathrm{tol}_k}$. In practice, the transmitter sets $\Gamma_{\mathrm{req}}\gg \Gamma_{\mathrm{tol}_k}>0,\forall k\in\{1,\ldots,K-1\}$, to ensure secure communication. Specifically, if the above optimization problem is feasible and  passive eavesdroppers do not exist in the system, the adopted problem formulation  guarantees  that the secrecy capacity is bounded below by $C_{\mathrm{sec}}\ge \log_2(1+\Gamma_{\mathrm{req}})-\log_2(1+\underset{k}{\max}\{\Gamma_{\mathrm{tol}_k}\})\ge 0$. We note that although $\Gamma_{\mathrm{req}}$ and $\Gamma_{\mathrm{tol}_k}$ in C1 and C2, respectively,  are
not optimization variables in this paper, a balance between
secrecy capacity and system capacity can be struck
by varying their values.   In C3, $\Gamma_{\mathrm{tol}}$ represents the maximum received SINR  tolerance  for successfully decoding at passive eavesdropper $j$. This constraint  specifies the minimum outage requirement at all passive eavesdroppers. In particular, the  maximum received SINR  among all passive eavesdroppers is required to be smaller than the maximum tolerable received SINR\footnote{In fact, the system operator can adjust  the values of the maximum received SINR  tolerance  $\Gamma_{\mathrm{tol}}$ to  account for performance variations due to potential system model mismatches such as the assumption that  $\mathbf{w}_{\mathrm{E}}$ is Gaussian distributed.} $\Gamma_{\mathrm{tol}}$   with at least probability $\kappa$.  For instance, if $\kappa=0.99$, $\Gamma_{\mathrm{req}}\ge\Gamma_{\mathrm{tol}}$,  and an idle receiver does not exist in the system, constraints  C1 and C3 together guarantee that the secrecy capacity between the transmitter and the desired receiver is bounded below by $C_{\mathrm{sec}}= \log_2(1+\Gamma_{\mathrm{req}})-\log_2(1+\underset{j}{\max}\big\{\Gamma_{\mathrm{PE}_j}\big\})\ge\log_2(1+\Gamma_{\mathrm{req}})-\log_2(1+\Gamma_{\mathrm{tol}})$ with  probability $0.99$.  We note that the proposed problem formulation and resource allocation schemes are also valid for $\kappa=0$. Specifically,  for $\kappa=0$ and continuous random variables $\Gamma_{\mathrm{PE}_j}\in(0,\infty)$, constraint C3 is always satisfied. Hence,  for $\kappa=0$, constraint C3 can be safely removed from the optimization problem formulation without loss of optimality.  Besides, although  the number of eavesdroppers  $J$ is not known at the transmitter, $J$ in C3 represents the maximum tolerable number of passive eavesdroppers that the transmitter can handle. Furthermore, we note that we do not  maximize the secrecy capacity in this paper as this would not necessarily lead to a power efficient resource allocation. $P_{\min}$ and $P_{\min_k}$ in C4 and C5 set the minimum required power transfer to the desired information  receiver and idle receiver $k$, respectively. We note that for given CSI uncertainty sets $\Omega_k,\forall k$, the transmitter can only guarantee the minimum required power transfer to the $K-1$ idle receivers if they use all their received power for energy harvesting, i.e., the idle receivers do not intend to eavesdrop.
$\eta$ in C4 and C5 denotes the energy harvesting efficiency of the receivers in converting
the received radio signal to electrical energy. Furthermore, in practice, each transmit antenna has its own power
amplifier in its analog front-end in practice, cf. Figure \ref{fig:system_model}. Hence, the  power radiated by each antenna is limited by the maximum transmit power of each power amplifier.  In C6, we take this physical limitation  into account by limiting the maximum transmit power from antenna $n$ to $P_{\max_n}$. We note that the introduction of a joint power consumption constraint for all power amplifiers is not necessary as long as $\sum_{n=1}^{N_{\mathrm{T}}} P_{\max_n}\le P_{\mathrm{PG}}$, where $P_{\mathrm{PG}}$ is the maximum power supply at the transmitter. C7 is the boundary constraint for the power splitting variable $\rho$. C8 and $\mathbf{V},\mathbf{W}_\mathrm{E}\in \mathbb{H}^{N_{\mathrm{T}}}$ constrain matrices $\mathbf{V}$ and $\mathbf{W}_{\mathrm{E}}$ to be positive semi-definite Hermitian matrices such that both are valid covariance matrices.

\begin{Remark}
We would like to emphasize that the considered problem formulation is a generalization of the cases where only artificial noise  or only  energy signal allocation are performed. If  $\mathbf{w}_{\mathrm{E}}$  is not known  at the legitimate receivers, it cannot be cancelled at the legitimate receivers. In this case,  we can set $\mathbf{w}_{\mathrm{E}}=\mathbf{0}$ in the problem formulation without loss of optimality. This is because if $\mathbf{w}_{\mathrm{E}}\ne\mathbf{0}$ and interference cancellation cannot be performed, we can define a new resource allocation policy with a new artificial noise signal, $\mathbf{v}_{\mathrm{new}}=\mathbf{v}+\mathbf{w}_{\mathrm{E}}$, and a new energy signal, $\mathbf{w}_{\mathrm{E}_{\mathrm{new}}}=\mathbf{0}$, such that the new resource allocation  policy  consumes the same amount of energy as the original solution and satisfies all the constraints.
\end{Remark}

\section{Solution of the Optimization Problem} \label{sect:solution}
The optimization problem in (\ref{eqn:cross-layer}) is a non-convex quadratically constrained quadratic program (QCQP) which involves chance constrained programming \cite{JR:chance_programming1,JR:chance_programming2} and  semi-infinite programming.  In particular, the non-convexity with respect to the beamforming vector $\mathbf{w}$ for the information signal and the power splitting ratio $\rho$ is due to
constraints C1 and C4. Besides, constraints C2 and C5 involve  infinitely many inequality constraints due to the continuity of the CSI uncertainty sets. Furthermore, the probabilistic (chance) constraint C3 couples all optimization variables and is non-convex.
In general, even if we remove chance constraint C3, there is no standard approach for solving non-convex optimization problems. In some extreme cases, an exhaustive search approach is required to obtain a globally optimal solution which is computationally intractable even for a moderate  system size. In order to derive an efficient
 resource allocation algorithm for the considered problem, we first recast the original problem as a semi-definite programming (SDP) problem in order to avoid the non-convexity associated with constraints C1 and C4. Then, we convert the infinite number of constraints in C2 and C5 into an equivalent finite number of  constraints. Subsequently, we propose a tractable convex constraint as a replacement for non-convex constraint C3. The reformulated problem with constraint replacement serves as a performance lower bound for the original problem formulation.  Finally, we use semi-definite programming relaxation (SDR) \cite{CN:SDP_relaxation1,JR:SDP_relaxation1} to obtain the optimal resource allocation solution for the reformulated problem. In practice,  the considered problem may be infeasible when the QoS requirements are stringent and/or the channels are in unfavourable  conditions. However,  in the sequel, we assume that the problem is always feasible for studying the design of different resource allocation schemes.

\subsection{Semi-definite Programming Relaxation} \label{sect:solution_dual_decomposition}
To facilitate SDP relaxation, we define $\mathbf{W}=\mathbf{w}\mathbf{w}^H$ and rewrite problem (\ref{eqn:cross-layer}) in terms of $\mathbf{W}$ as
\begin{eqnarray}
\label{eqn:SDP}&&\hspace*{-5mm} \underset{\mathbf{W,V,}\mathbf{W}_\mathrm{E}\in \mathbb{H}^{N_{\mathrm{T}}}, \rho
}{\mino}\,\, \Tr(\mathbf{W})+\Tr(\mathbf{V})+\Tr(\mathbf{W}_\mathrm{E})\nonumber\\
\notag \hspace*{-1mm}\mbox{s.t.} &&\hspace*{-6mm}\mbox{C1: }\notag\frac{\Tr(\mathbf{H}\mathbf{W})}
{\sigma_{\mathrm{ant}}^2+\Tr(\mathbf{H}\mathbf{V})+\frac{\sigma_{\mathrm{s}}^2}{\rho} } \ge \Gamma_{\mathrm{req}}, \\
&&\hspace*{-7mm}\mbox{C2: }\max_{\Delta\mathbf{g}_k\in {\Omega}_k}\,\,\notag\frac{\Tr(\mathbf{G}_k\mathbf{W})}{\sigma_{\mathrm{ant}}^2
+\Tr(\mathbf{G}_k\mathbf{V})+\sigma_{\mathrm{s}}^2 } \le \Gamma_{\mathrm{tol}_k},\forall k, \\
&&\hspace*{-7mm}\mbox{C3: }\notag\Pr\Big(\hspace*{-0.5mm}\max_{j\in\{1,\ldots,J\}} \Big\{\frac{\Tr(\mathbf{\tilde{L}}_j\mathbf{W})}{\Tr(\mathbf{\tilde{L}}_j\mathbf{W}_\mathrm{E})+
\Tr(\mathbf{\tilde{L}}_j\mathbf{V})+\tilde\sigma^2_j }\Big\}\hspace*{-0.5mm} \le\hspace*{-0.5mm} \Gamma_{\mathrm{tol}}\hspace*{-0.5mm}\Big)\notag\\
&&\hspace*{1mm}\notag\ge \kappa,\\
&&\hspace*{-7mm}\mbox{C4: }\notag \Tr(\mathbf{H}\mathbf{W})+\Tr(\mathbf{H}\mathbf{W}_\mathrm{E})+\Tr(\mathbf{H}\mathbf{V})\\
&&\hspace*{1mm}\notag+\sigma_{\mathrm{ant}}^2\ge\frac{ P_{\min}}{(1-\rho)\eta}, \\
&&\hspace*{-7mm}\mbox{C5: }\notag\min_{\Delta\mathbf{g}_k\in {\Omega}_k}\,\,\Tr(\mathbf{G}_k\mathbf{W})+\Tr(\mathbf{G}_k\mathbf{V})+\Tr(\mathbf{G}_k\mathbf{W}_\mathrm{E})\notag\\
&&\hspace*{1mm}\notag+\sigma_{\mathrm{ant}}^2\ge \frac{P_{\min_k}}{\eta},\forall k, \\
&&\hspace*{-7mm}\mbox{C6: }\notag\hspace*{-0.5mm} \Tr\Big(\hspace*{-0.5mm}\mathbf{\Psi}_n\big(\mathbf{W}\hspace*{-0.5mm} +\hspace*{-0.5mm} \mathbf{V}\hspace*{-0.5mm} +\hspace*{-0.5mm} \mathbf{W}_{\mathrm{E}}\big)\hspace*{-0.5mm}\Big)\le P_{\max_n},\, \forall n\hspace*{-0.5mm}\in\hspace*{-0.5mm}\{1,\ldots,N_{\mathrm{T}}\}, \\
&&\hspace*{-7mm}\mbox{C7:}\,\, 0\le\rho\le 1, \,\,\,\mbox{C8:}\,\, \mathbf{W}\succeq \mathbf{0}, \mathbf{V}\succeq \mathbf{0},\mathbf{W}_{\mathrm{E}}\succeq \mathbf{0},\notag \\
&&\hspace*{-7mm}\mbox{C9:}\,\, \Rank(\mathbf{W})=1,
\end{eqnarray}
where $\mathbf{W}\succeq \mathbf{0}$, $\mathbf{W}\in \mathbb{H}^{N_{\mathrm{T}}}$, and $\Rank(\mathbf{W})=1$ in (\ref{eqn:SDP}) are imposed to guarantee that $\mathbf{W}=\mathbf{w}\mathbf{w}^H$ holds after optimizing $\mathbf{W}$. We note that the  per-antenna radiated power in constraint C6 in (\ref{eqn:cross-layer}) can be represented as $\Tr\Big(\mathbf{\Psi}_n\big(\mathbf{W}+\mathbf{V}+\mathbf{W}_{\mathrm{E}}\big)\Big)$, where $\mathbf{\Psi}_n=\mathbf{e}_n\mathbf{e}_n^H$ and $\mathbf{e}_n\in\mathbb{R}^{N_{\mathrm{T}}\times 1}$ is the $n$-th unit vector  of length $N_{\mathrm{T}}$, i.e., $\big[\mathbf{e}_n\big]_{n,1}=1$ and $\big[\mathbf{e}_n\big]_{b,1}=0,\forall b\ne n,b\in\{1,\ldots,N_{\mathrm{T}}\}$. After variable transformation $\mathbf{W}=\mathbf{w}\mathbf{w}^H$ and some manipulations, it can be observed that constraints C1 and C4 are convex with respect to $\{\mathbf{W,V,}\mathbf{W}_\mathrm{E}, \rho\}$.  Next, we handle constraints C2 and C5. Although constraints C2 and C5 are convex with respect to the optimization variables (after some mathematical manipulations), they are semi-infinite constraints  which are generally intractable  for resource allocation algorithm design. To facilitate the  solution, we transform constraints C2 and C5 into linear matrix inequalities (LMIs) using the following lemma:
\begin{Lem}[S-Procedure \cite{book:convex}] Let a function $f_m(\mathbf{x}),m\in\{1,2\},\mathbf{x}\in \mathbb{C}^{N\times 1},$ be defined as
\begin{eqnarray}
f_m(\mathbf{x})=\mathbf{x}^H\mathbf{A}_m\mathbf{x}+2 \mathrm{Re} \{\mathbf{b}_m^H\mathbf{x}\}+c_m,
\end{eqnarray}
where $\mathbf{A}_m\in\mathbb{H}^N$, $\mathbf{b}_m\in\mathbb{C}^{N\times 1}$, and $c_m\in\mathbb{R}^{1\times 1}$. Then, the implication $f_1(\mathbf{x})\le 0\Rightarrow f_2(\mathbf{x})\le 0$  holds if and only if there exists a $\delta\ge 0$ such that
\begin{eqnarray}\delta
\begin{bmatrix}
       \mathbf{A}_1 & \mathbf{b}_1          \\
       \mathbf{b}_1^H & c_1           \\
           \end{bmatrix} -\begin{bmatrix}
       \mathbf{A}_2 & \mathbf{b}_2          \\
       \mathbf{b}_2^H & c_2           \\
           \end{bmatrix}          \succeq \mathbf{0}         ,
\end{eqnarray}
provided that there exists a point $\mathbf{\hat{x}}$ such that $f_k(\mathbf{\hat{x}})<0$.
\end{Lem}

As a result, we can apply Lemma 1 to constraint C2. In particular, we substitute
 $\mathbf{g}_k=\mathbf{\hat g}_k +\Delta\mathbf{g}_k$ in constraint C2. Therefore, the implication
\begin{eqnarray}
 &&\Delta\mathbf{g}_k^H \Delta\mathbf{g}_k\le\varepsilon_k^2\\
\Rightarrow\hspace*{-6mm}&&0\ge \Delta\mathbf{g}_k^H\big(\frac{\mathbf{W}}{\Gamma_{\mathrm{tol}_k}}\hspace*{-0.5mm}-\hspace*{-0.5mm}\mathbf{V}\big)\Delta\mathbf{g}_k\hspace*{-0.5mm}+\hspace*{-0.5mm}2\mathrm{Re}\Big\{\hspace*{-0.5mm}\mathbf{\hat g}_k^H\big(\frac{\mathbf{W}}{\Gamma_{\mathrm{tol}_k}}-\mathbf{V}\big)\Delta\mathbf{ g}_k\hspace*{-0.5mm}\Big\}\notag\\
&&+
\mathbf{\hat g}_k^H\big(\frac{\mathbf{W}}{\Gamma_{\mathrm{tol}_k}}\hspace*{-0.5mm}-\hspace*{-0.5mm}\mathbf{V}\big)\mathbf{\hat g}_k\hspace*{-0.5mm}-\hspace*{-0.5mm}\sigma_{\mathrm{ant}}^2\hspace*{-0.5mm}-\hspace*{-0.5mm}\sigma_{\mathrm{s}}^2,\forall k,\notag
\end{eqnarray}
holds if and only if there exist $\delta_k\ge 0, k\in\{1,\ldots,K-1\},$  such that the following  LMI constraints hold:
\begin{eqnarray}\label{eqn:LMI_C2}
&&\mbox{C2: }\mathbf{S}_{\mathrm{C}_{2_k}}\big(\mathbf{W},\mathbf{V},\delta_k\big)\notag\\
=&&
          \begin{bmatrix}
       \delta_k\mathbf{I}_{N_{\mathrm{T}}}+\mathbf{V} & \mathbf{V}\mathbf{\hat g}_k          \\
       \mathbf{\hat g}_k^H \mathbf{V}    & -\delta_k\varepsilon_k^2 +\sigma_{\mathrm{ant}}^2 +\sigma_{\mathrm{s}}^2+  \mathbf{\hat g}_k^H \mathbf{V} \mathbf{\hat g}_k        \\
           \end{bmatrix}\notag\\
            -&&\frac{1}{\Gamma_{\mathrm{tol}_k}} \mathbf{U}_{\mathbf{g}_k}^H\mathbf{W}\mathbf{U}_{\mathbf{g}_k}\succeq \mathbf{0}, \forall k,
\end{eqnarray}
where $\mathbf{U}_{\mathbf{g}_k}=\Big[\mathbf{I}_{N_{\mathrm{T}}}\quad\mathbf{\hat g}_k\Big]$. Similarly, by using Lemma 1, constraint C5 can be equivalently written as
\begin{eqnarray}\label{eqn:LMI_C5}&&\mbox{C5: }\mathbf{S}_{\mathrm{C}_{5_k}}(\mathbf{W},\mathbf{V},\mathbf{W}_\mathrm{E},\nu_k)\\
=&&\notag
           \begin{bmatrix}
       \nu_k\mathbf{I}_{N_{\mathrm{T}}}+\mathbf{V}& \mathbf{V}\mathbf{\hat g}_k          \\
       \mathbf{\hat g}_k^H \mathbf{V}
        & -\nu_k\varepsilon_k^2 -\frac{P_{\min}}{\eta} +\sigma_{\mathrm{ant}}^2 +  \mathbf{\hat g}_k^H \mathbf{V} \mathbf{\hat g}_k        \\
           \end{bmatrix}\notag\\
           +&& \mathbf{U}_{\mathbf{g}_k}^H\mathbf{W}_\mathrm{E}\mathbf{U}_{\mathbf{g}_k}+ \mathbf{U}_{\mathbf{g}_k}^H\mathbf{W}\mathbf{U}_{\mathbf{g}_k}\succeq \mathbf{0}, \forall k,\notag
\end{eqnarray}
for $\nu_k\ge 0, k\in\{1,\ldots,K-1\}$. We note that now constraints C2 and C5 involve only a finite number of constraints which facilitates the resource allocation algorithm design.

The third obstacle in solving (\ref{eqn:SDP}) is the probabilistic (chance) constraint C3 and the problem formulation is known as chance constrained programming.  In general, constraint C3 is non-convex  since the optimization variables are coupled. To overcome this problem, we introduce the following lemma.
\begin{Lem}\label{lemma:chance_constraint} Assuming the normalized upper bound channel gain vectors of the passive eavesdroppers, $\mathbf{\tilde L}_j,j\in\{1,\ldots,J\}$, can be modeled as independent and identical distributed (i.i.d.) Rayleigh random variables\footnote{We assume that the passive eavesdroppers are not collaborating  and are physically separated by  a distance of at least half a wavelength. For a carrier frequency of $470$ MHz, half of a wavelength is roughly equal to $30$ cm which is a reasonable assumption for the minimum receiver separation in practice. Besides, we assume that there is a  sufficient number of scatterers in the channels between the transmitter and the passive eavesdroppers such that these channels can be modelled as Rayleigh distribution due to the central limit theorem \cite{book:david_wirelss_com}.   These assumptions are commonly used in the literature \cite{JR:AN_MISO_secrecy,JR:independent_channel} when studying resource allocation algorithm design for providing physical layer security.
},  the following constraint implication holds:
\begin{eqnarray}\label{eqn:chance_constraint}
 \notag\overline{\mbox{C3}}\mbox{: }&&\Big(\Phi^{-1}_{N_{\mathrm{T}}}(1-\kappa^{1/J})\Gamma_{\mathrm{tol}}\tilde\sigma^2\Big)\mathbf{I}_{\mathrm{N_T}}\\
\succeq &&\mathbf{W}\hspace*{-0.5mm}-\hspace*{-0.5mm}\Gamma_{\mathrm{tol}}\mathbf{W}_\mathrm{E}
\hspace*{-0.5mm}-\hspace*{-0.5mm}\Gamma_{\mathrm{tol}}\mathbf{V}\\
\Longrightarrow&&{\mbox{C3: }}\Pr\Big(\max_{j\in\{1,\ldots,J\}} \Big\{\Gamma_{\mathrm{PE},j}\Big\} \le \Gamma_{\mathrm{tol}}\Big)\ge \kappa,
\end{eqnarray}
\end{Lem}
where $\Phi^{-1}_{N_{\mathrm{T}}}(\cdot)$ denotes the inverse
cumulative distribution function (c.d.f.) of an inverse central
chi-square random variable with 2$N_\mathrm{T}$ degrees of freedom.

\emph{\quad Proof: }Please refer to Appendix A for a proof of
Lemma \ref{lemma:chance_constraint}. \qed

As a result, we can replace\footnote{
 We note that a resource allocation policy which satisfies  constraint $\overline{\mbox{C3}}$ also satisfies ${\mbox{C3}}$, but not vice versa.  In other words, replacing constraint   ${\mbox{C3}}$ with $\overline{\mbox{C3}}$ results in a smaller feasible solution set for optimization which leads to a lower bound system performance. To quantify the impact of the constraint restriction, the resulting performance loss is quantified in  Section  \Rmnum{5} via simulations.
} the chance constraint C3 with the constraint in (\ref{eqn:chance_constraint}). Constraint $\overline{\mbox{C3}}$ is tractable in the sense that: (i) a feasible solution point satisfying  (\ref{eqn:chance_constraint}) is also feasible for C3, and (ii) the new constraint in (\ref{eqn:chance_constraint}) is convex with respect to the optimization variables \cite{book:convex}. Thus, substituting (\ref{eqn:LMI_C2}) and (\ref{eqn:LMI_C5}) into (\ref{eqn:SDP}) and replacing  constraint C3 by (\ref{eqn:chance_constraint}), we obtain the following optimization problem:
\begin{eqnarray}
\label{eqn:SDP-robust}&&\hspace*{-3mm} \underset{\mathbf{W,V,}\mathbf{W}_\mathrm{E}\in \mathbb{H}^{N_{\mathrm{T}}}, \rho,\boldsymbol \delta, \boldsymbol \nu
}{\mino}\,\, \Tr(\mathbf{W})+\Tr(\mathbf{V})+\Tr(\mathbf{W}_\mathrm{E})\nonumber\\
\notag \mbox{s.t.} &&\hspace*{8mm}\mbox{C1, C4, C6, C7, C8,} \\
&&\hspace*{-5mm}\mbox{C2: }\notag\mathbf{S}_{\mathrm{C}_{2_k}}(\mathbf{W},\mathbf{V},\delta_k)\succeq \mathbf{0},\forall k, \\
&&\hspace*{-5mm}\overline{\mbox{C3}}\mbox{: }\notag {\Big(\hspace*{-0.5mm}\Phi^{-1}_{N_{\mathrm{T}}}(1\hspace*{-0.5mm}-\hspace*{-0.5mm}\kappa^{1/J})\Gamma_{\mathrm{tol}}\tilde\sigma^2\hspace*{-0.5mm}\Big)
\mathbf{I}_{\mathrm{N_T}}\hspace*{-0.5mm}\succeq\hspace*{-0.5mm} \mathbf{W}\hspace*{-0.5mm}-\hspace*{-0.5mm}\Gamma_{\mathrm{tol}}\mathbf{W}_\mathrm{E}
\hspace*{-0.5mm}-\hspace*{-0.5mm}\Gamma_{\mathrm{tol}}\mathbf{V}}, \\
&&\hspace*{-5mm}\mbox{C5: } \notag\mathbf{S}_{\mathrm{C}_{5_k}}(\mathbf{W},\mathbf{V},\mathbf{W}_\mathrm{E},\nu_k)\succeq \mathbf{0},\forall k, \\
&&\hspace*{-5mm}\mbox{C9:}\,\, \Rank(\mathbf{W})=1,\quad\mbox{C10: }\delta_k,\nu_k \ge 0,\forall k,
\end{eqnarray}
where $\boldsymbol \delta$ and $\boldsymbol \nu$ are auxiliary  optimization variable vectors, whose elements $\delta_k\ge0, k\in\{1,\ldots,K-1\}$, and $\nu_k\ge0, k\in\{1,\ldots,K-1\}$, were introduced in (\ref{eqn:LMI_C2}) and (\ref{eqn:LMI_C5}), respectively.  Now, $\mbox{C9:} \Rank($  $\mathbf{W})  =1$ is  the remaining obstacle in solving the problem in (\ref{eqn:SDP-robust}).
By relaxing constraint $\mbox{C9: }\Rank(\mathbf{W})=1$, i.e., removing it from the problem formulation, the considered problem becomes a convex SDP which can be solved efficiently by numerical solvers such as SeDuMi \cite{JR:SeDumi} and  SDPT3 \cite{JR:SDPT3}. From the basic principles of optimization theory, if the obtained solution $\mathbf{W}$ for the relaxed problem admits a rank-one matrix, then it is the optimal solution of the original problem in (\ref{eqn:SDP-robust}). Then, the optimal $\mathbf{w}$ can be obtained by performing eigenvalue decomposition on $\mathbf{W}$. However, it is known that the constraint relaxation may not be tight and $\Rank(\mathbf{W})>1$ may occur.  In the following, we will first reveal a sufficient condition for $\Rank(\mathbf{W})=1$ for the relaxed problem. Then, we propose a method for constructing an optimal solution for the relaxed version of   (\ref{eqn:SDP-robust}) with a rank-one matrix $\mathbf{W}$. Furthermore, we  exploit the sufficient condition of the optimal solution and use it as a
building block for the design of a suboptimal resource allocation algorithm, which requires a lower computational complexity than the construction of the rank-one solution.

 \subsection{Optimality Conditions for SDP Relaxation}
In this subsection, we first reveal the tightness of the proposed  SDP relaxation. Then, we study a sufficient condition   for obtaining an optimal solution  which will be used to design  a low complexity suboptimal resource allocation scheme.

{ First, we introduce a theorem which reveals  the tightness of the SDP relaxation. This theorem  closely follows a similar theorem in \cite[Proposition 4.1]{JR:rui_zhang}\footnote{We note that the theorem  in  \cite[Proposition 4.1]{JR:rui_zhang} considers  a  communication system with separated receivers for  the case of perfect CSI and without passive eavesdroppers. In fact, the results of \cite{JR:rui_zhang} cannot be directly applied to the system model considered in this paper without the steps and transformations introduced in Sections \Rmnum{2} to \Rmnum{4}-A. }. }

\begin{Thm}\label{prop:construction} Suppose the optimal solution of the relaxed version of (\ref{eqn:SDP-robust}) is denoted as $\{\mathbf{W^*,V^*,}\mathbf{W}_\mathrm{E}^*, \rho^*$, $\boldsymbol \delta^*, \boldsymbol \nu^*\},{\Gamma}_{\mathrm{req}}>0$,  and  $\Rank(\mathbf{W}^*)>1$. Then, there exists a feasible solution of (\ref{eqn:SDP-robust}), denoted as  $\{\mathbf{\widetilde W^*},\mathbf{\widetilde V^*,} $ $ \mathbf{\widetilde W}_\mathrm{E}^*,  \widetilde \rho^*,\boldsymbol{\widetilde \delta}^*,  \boldsymbol{\widetilde \nu}^*\}$, which not only achieves the same objective value as $\{\mathbf{W^*,V^*,}\mathbf{W}_\mathrm{E}^*, \rho^*,\boldsymbol \delta^*, \boldsymbol \nu^*\}$, but also admits a rank-one matrix $\mathbf{\widetilde W}^*$, i.e.,  $\Rank(\mathbf{\widetilde W}^*)=1$.
\end{Thm}
\emph{\quad Proof: } Please refer to Appendix B for a proof of Theorem \ref{prop:construction} and the method for constructing $\{\mathbf{\widetilde W^*},\mathbf{\widetilde V^*,}  \mathbf{\widetilde W}_\mathrm{E}^*,$ $ \widetilde \rho^*, \boldsymbol{\widetilde \delta}^*,  \boldsymbol{\widetilde \nu}^*\}$ with $\Rank(\mathbf{\widetilde W^*})=1$.\qed

As discussed  in Appendix B, constructing an optimal solution set $\{\mathbf{\widetilde W^*},\mathbf{\widetilde V^*,}  \mathbf{\widetilde W}_\mathrm{E}^*, \widetilde \rho^*,\boldsymbol{\widetilde \delta}^*,  \boldsymbol{\widetilde \nu}^*\}$ with  $\Rank($ $ \mathbf{\widetilde W^*})=1$  may require the solution of the dual problem and one extra optimization problem, cf. Remark \ref{Remark:construction}.  Besides, knowledge of the Lagrange multiplier matrix $\mathbf{Y}^*$ is required for constructing the rank-one solution which may not be provided by some numerical solvers. Furthermore, the associated signal processing delay and computational complexity may not be affordable when the transmitter has limited signal processing capability. Thus, in the following,  we  present a sufficient  condition for obtaining a rank-one matrix solution $\mathbf{W}$ for the relaxed version of problem (\ref{eqn:SDP-robust}) which will also be used for  designing a low complexity suboptimal resource allocation algorithm.
\begin{proposition}\label{prop:sufficient} Consider the relaxed version  of problem (\ref{eqn:SDP-robust}) for $\Gamma_{\mathrm{req}}>0$ and denote $\mathbf{D}^*_{\mathrm{C}_{2_k}}$ and $\mathbf{D}^*_{\mathrm{C}_{5_k}}$  as the optimal Lagrange multiplier matrices associated with constraints C2 and C5, respectively. A sufficient condition for $\Rank(\mathbf{W}^*)=1$ is that  $\frac{\mathbf{D}^*_{\mathrm{C}_{2_k}}}{\Gamma_{\mathrm{tol}_k}}-\mathbf{D}^*_{\mathrm{C}_{5_k}}\succeq \mathbf{0}$, $\forall k\in\{1,\ldots,K-1\}$, holds.
\end{proposition}
\emph{\quad Proof: } Please refer to Appendix C for a  proof of Proposition \ref{prop:sufficient}. \qed

Intuitively, when the requirement in constraint C5 becomes less stringent, i.e., $P_{\min_k}\rightarrow 0$, the corresponding Lagrange multiplier matrix   $ \mathbf{D}^*_{\mathrm{C}_{5_k}}\to \mathbf{0}$ since the constraint is less often active. As a result,  the  SDP relaxation algorithm has a higher chance of obtaining a rank-one matrix solution and thus achieves the global optimum with respect to the reformulated problem.

\begin{Remark}
We would like to emphasize that although Proposition 1 provides a sufficient condition for a rank-one matrix solution under SDP relaxation, we found by simulation that the proposed SDP relaxation may result in a rank-one matrix $\mathbf{W}^*$ even if the sufficient condition does not hold.
\end{Remark}

\subsubsection*{Suboptimal resource allocation scheme}
  According to Proposition 1, the solution of (\ref{eqn:SDP-robust}) has a rank-one $\mathbf{W}^*$ when constraint C5 is not active, i.e., $\mathbf{D}^*_{\mathrm{C}_{5_k}}=\mathbf{0},\forall k$, or constraint C5 is independent of optimization variable $\mathbf{W}$. For facilitating an efficient resource allocation algorithm design, we replace constraint  C5 in (\ref{eqn:SDP-robust}) by $\overline{\mbox{C5}}$ and the new optimization problem becomes
\begin{eqnarray}\label{eqn:suboptimal1}\notag
&&\hspace*{1mm} \underset{\mathbf{W,V,}\mathbf{W}_\mathrm{E}\in \mathbb{H}^{N_{\mathrm{T}}}, \rho,\boldsymbol \delta, \boldsymbol \nu
}{\mino}\,\, \Tr(\mathbf{W})+\Tr(\mathbf{V})+\Tr(\mathbf{W}_\mathrm{E})\\
\notag \mbox{s.t.} &&\hspace*{10mm}\mbox{C1, C2, } \overline{\mbox{C3}},\mbox{ C4 -- C8, C10, }\\
&&\hspace*{15mm}\notag\overline{\mbox{C5}}\mbox{: }\mathbf{\tilde S}_{\mathrm{C}_{5_k}}(\mathbf{V},\mathbf{W}_\mathrm{E},\nu_k)\\
=&& \begin{bmatrix}
       \nu_k\mathbf{I}_{N_{\mathrm{T}}}+\mathbf{V}& \mathbf{V}\mathbf{\hat g}_k          \\
       \mathbf{\hat g}_k^H \mathbf{V}
        & -\nu_k\varepsilon_k^2 -\frac{P_{\min}}{\eta} +\sigma_{\mathrm{ant}}^2 +  \mathbf{\hat g}_k^H \mathbf{V} \mathbf{\hat g}_k        \\
           \end{bmatrix}\notag\\
          +&&  \mathbf{U}_{\mathbf{g}_k}^H\mathbf{W}_\mathrm{E}\mathbf{U}_{\mathbf{g}_k}\succeq \mathbf{0}, \forall k.
\end{eqnarray}
Compared to constraint C5, the new constraint $\overline{\mbox{C5}}$ reduces the feasible solution set for (\ref{eqn:SDP-robust}) as the contribution of the information signal to the energy harvesting, i.e., $\mathbf{W}$, at  idle receiver $k$  is neglected. Thus, the obtained solution of problem (\ref{eqn:suboptimal1}) serves as a performance lower bound for the reformulated optimization problem (\ref{eqn:SDP-robust}). We note that the new constraint preserves the convexity of the relaxed problem and (\ref{eqn:suboptimal1}) can be solved efficiently via SDP relaxation and the aforementioned  numerical solvers. Furthermore, it can be shown that the obtained solution\footnote{We can follow a similar approach as in Appendix C to examine the KKT conditions for the new problem formulation in (\ref{eqn:suboptimal1}). In particular,  the sufficient condition for a rank-one matrix solution stated in Proposition 1 is always satisfied for the new problem formulation since  constraint $\overline{\mbox{C5}}$ is independent of $\mathbf{W}$.} of problem (\ref{eqn:suboptimal1}) has always a rank-one beamforming matrix, i.e., $\Rank(\mathbf{W})=1$, even though SDP relaxation is applied.
%

\begin{figure*}[t]
 \centering
\includegraphics[angle=-90,scale=0.5]{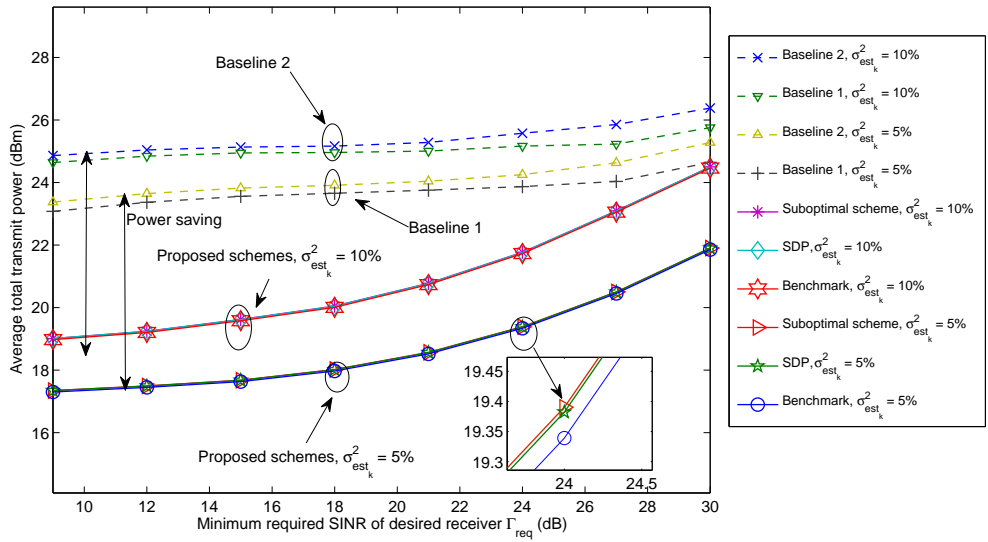}\vspace*{-4mm}
\caption{Average total transmit power (dBm) versus the minimum required SINR of the desired receiver, $\Gamma_{\mathrm{req}}$ (dB),  for different resource allocation schemes.
The double-sided arrows indicate the power savings achieved by the proposed schemes compared to the baseline schemes.} \label{fig:p_SNR}
\end{figure*}

\section{Results}
\label{sect:result-discussion}In this section, we evaluate the
system performance for the proposed resource allocation schemes using simulations.  The TGn path loss model \cite{report:tgn} for indoor communications is adopted with a carrier
center frequency of $470$ MHz \cite{report:80211af}.  The directional joint transmit and receive antenna gain is  $10$ dB. The reference distance of the path loss model is $2$ meters  and there are $K$ receivers uniformly distributed between
the reference distance and the maximum service distance of $20$ meters. The wavelength of the carrier signal is $0.6$ meter which is smaller than the minimum distance between the transmitter and the receivers. Thus, the far-field  assumption for the channel model in \cite{report:tgn} holds.  The transmitter is equipped with $N_{\mathrm{T}}$ antennas. The small scale fading coefficients between the transmitter and the legitimate receivers
are generated as independent and identically distributed Rician random
variables with Rician factor $6$ dB. The corresponding shadowing effect is assumed to be $0$ dB. On the other hand, we assume that  there are $J=5$ eavesdroppers eavesdropping the information from outdoors. Thus,  the multipath fading coefficients between the transmitter and the $J$ passive eavesdroppers  are modeled as Rayleigh random variables.  We set  $\kappa=0.99$ for providing secure communication.
  We assume that  the signal processing noise power\footnote{We assume that the signal processing noise power at each receiver is due to quantization noise in the analog-to-digital (ADC) converter. Specifically, a 10-bit uniform quantizer is used for quantizing the received signal at the receivers. } and the antenna
noise power are $\sigma_{\mathrm{s}}^2=-35$ dBm
and $\sigma_{\mathrm{ant}}^2=-111$ dBm, respectively.  Unless specified otherwise, we assume  a minimum required power transfer of $P_{\min}=P_{\min_k}=0$ dBm, $\forall k$, and an energy harvesting efficiency of $\eta=0.5$.  On the other hand,  the maximum  SINR tolerance of each idle receiver (potential eavesdropper) and passive eavesdropper  is set to  $\Gamma_{\mathrm{tol}_k}=\Gamma_{\mathrm{tol}}=0$ dB.  The maximum transmit power
per-antenna is set to  $P_{\max_n}=30$ dBm, $\forall n\in\{1,\ldots,N_{\mathrm{T}}\}$. To facilitate the presentation in the sequel, we define the normalized maximum  channel estimation error of idle receiver $k$  as  $\sigma_{\mathrm{est}_k}^2=\frac{\varepsilon^2_k}{\norm{\mathbf{g}_k}^2}$ where $\sigma_{\mathrm{est}_a}^2=\sigma_{\mathrm{est}_b}^2,\forall a, b\in\{1,\ldots,K-1\}$. The average system performance shown in the following sections is obtained by averaging over different realizations of  both path loss and multipath  fading.

\subsection{Average Total Transmit Power versus Minimum Required SINR}
Figure \ref{fig:p_SNR} depicts the  average total transmit power versus the
minimum required SINR of the desired receiver, $\Gamma_{\mathrm{req}}$, for $K=4$ legitimate receivers,  different resource allocation schemes, and different normalized maximum channel estimation errors, $\sigma_{\mathrm{est}_k}^2$. The transmitter has $N_{\mathrm{T}}=6$ transmit antennas.
It can be observed that the average total transmit power of the proposed SDP resource allocation scheme is a monotonically non-decreasing function of $\Gamma_{\mathrm{req}}$. This is attributed to the fact that a higher transmit power is necessary for satisfying constraint C1 when the minimum SINR requirement of $\Gamma_{\mathrm{req}}$ becomes more stringent. Besides, it can be seen that the average total transmit power increases for increasing normalized maximum  channel estimation errors, $\sigma_{\mathrm{est}_k}^2$. In fact, with  increasing imperfectness of the channel estimation, the transmitter has to allocate more power to the artificial noise and the energy signal to prevent  interception by potential eavesdroppers and to facilitate efficient energy transfer for fulling constraints C2 and C5, respectively. { Furthermore, we  investigate the performance loss incurred by replacing the non-convex probabilistic constraint C3 with the proposed convex deterministic constraint $\overline{\mbox{C3}}$, cf. Lemma 2 and (\ref{eqn:SDP-robust}), in Figure \ref{fig:p_SNR}.} In particular,  we compare the performance of the  proposed optimal SDP resource allocation scheme obtained based on (\ref{eqn:SDP-robust})  with that of a benchmark system. Specifically,  the performance of the benchmark system is computed by solving (\ref{eqn:cross-layer}) for $\kappa=0$ in constraint C3. In other words, we assume that the passive eavesdropper does not exist in the system. Thus, the performance gap between the curves of the SDP resource allocation scheme and the benchmark system  constitutes an upper bound on the performance loss incurred by the constraint replacement.   Figure \ref{fig:p_SNR} shows that for a wide range of minimum required SINRs of the desired receiver,  the performance loss  is less than $0.1$ dB, despite the fact that  the
reformulated problem  guarantees  communication security robustness against passive eavesdropping.

For comparison, Figure
\ref{fig:p_SNR} also contains the average total transmit power of the  proposed suboptimal scheme and two baseline
resource  allocation schemes.
For baseline scheme 1, we adopt an isotropic radiation pattern for $\mathbf{W}_{\mathrm{E}}$ as the CSI of the passive eavesdroppers is not known at the transmitter \cite{JR:Artifical_Noise1}. Then,  we minimize the total transmitted power by optimizing   $\mathbf{W},\mathbf{V},\rho,\boldsymbol{\delta},  \boldsymbol{\nu}$, and the power allocated to $\mathbf{w}_{\mathrm{E}}$ subject to the same constraints as in (\ref{eqn:SDP-robust}) via SDP relaxation. For baseline scheme 2, we adopt maximum ratio transmission (MRT) \cite{CN:Kwan_PIMRC2013} for delivering  the information signal to the desired legitimate receiver. Similar to baseline scheme 1,  an isotropic radiation pattern is adopted for  the energy signal. Then, we optimize the power allocated to $\mathbf{W}_{\mathrm{E}}$, the MRT beamforming matrix $\mathbf{W}$, $\mathbf{V}$,  $\rho,\boldsymbol{\delta}$, and $\boldsymbol{\nu}$ for minimization of the total transmit power subject to the same constraints as in (\ref{eqn:SDP-robust}).
It can be seen that the  proposed suboptimal scheme  closely approaches the performance of the benchmark system and the performance achieved by SDP relaxation of the reformulated problem, despite the fact that  the proposed suboptimal scheme employs  a smaller feasible solution set compared to the original problem formulation in (\ref{eqn:cross-layer}). On the other hand, it can be observed that the lower computational complexity of the baseline schemes comes at the expense of a significantly  higher transmit power compared to the other schemes. Indeed, the proposed SDP resource allocation scheme and the suboptimal scheme fully utilize the CSI of all communication links and  optimize the space spanned by  the energy signal and the artificial noise for performing resource allocation. On the contrary, for the baseline schemes,  the transmitter is unable to fully exploit the available degrees of freedom in resource allocation
since the energy signal is radiated  isotropically and/or  $\mathbf{W}$ is fixed. Yet, even though the available CSI at the transmitter is imperfect, all the schemes (including the baseline schemes) are able to fulfill the QoS requirements due to the  proposed robust  resource allocation algorithm design and the optimization of the artificial noise covariance matrix $\mathbf{V}$.

\begin{figure}[t]\hspace*{-4mm}
\subfigure[Normalized maximum channel estimation errors $\sigma_{\mathrm{est}_k}^2=5\%$.]{\centering
\includegraphics[scale=0.5]{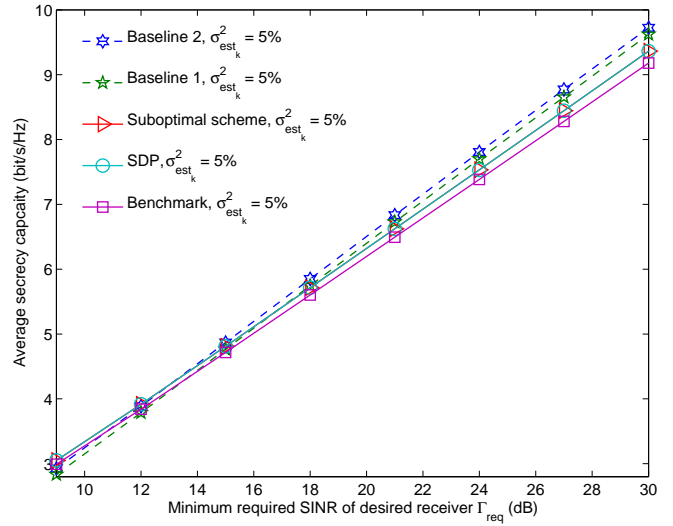}
\label{fig:fig:cap_SNR_a} }  \subfigure[Normalized maximum channel estimation errors $\sigma_{\mathrm{est}_k}^2=10\%$.]{\centering\hspace*{-4mm}
\includegraphics[scale=0.5]{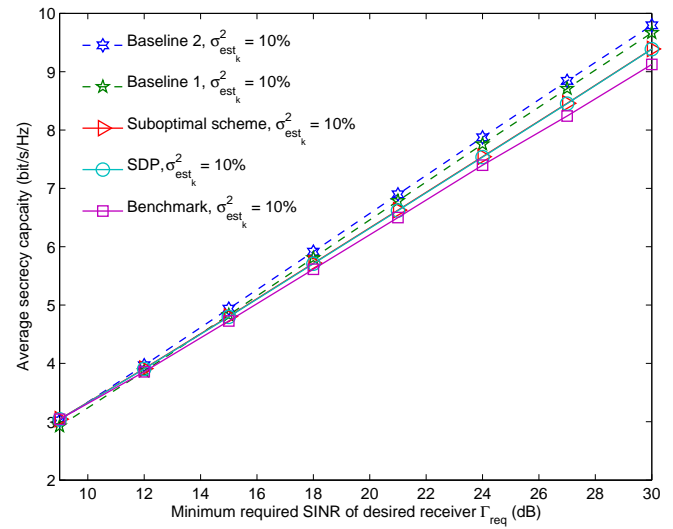}
\label{fig:fig:cap_SNR_b} }\caption[Optional caption for list of figures]
{Average system secrecy capacity (bit/s/Hz) versus the minimum required SINR of the desired receiver,  $\Gamma_{\mathrm{req}}$ (dB), for different  normalized maximum channel estimation errors $\sigma_{\mathrm{est}_k}^2$, and different resource allocation schemes.}\label{fig:cap_SNR}
\end{figure}

Figure \ref{fig:fig:cap_SNR_a} and Figure \ref{fig:fig:cap_SNR_b} illustrate the average system secrecy capacity versus the minimum required SINR of the desired receiver, $\Gamma_{\mathrm{req}}$, for $K=4$ receivers, $N_{\mathrm{T}}=6$ transmit antennas,  different resource allocation schemes, and different  normalized maximum channel estimation errors, $\sigma_{\mathrm{est}_k}^2$.  It can be seen that the average system secrecy capacity, i.e., $C_{\mathrm{sec}}$, increases with  $\Gamma_{\mathrm{req}}$
since the maximum tolerable SINRs of the idle receivers are constrained to $\Gamma_{\mathrm{tol}_k}=0$ dB. Besides, as expected,  the baseline schemes achieve on average a higher  secrecy capacity in the high transmit power regime compared to the other schemes. However, the superior performance in terms of secrecy capacity comes at the expense of an exceedingly high transmit power, cf. Figure \ref{fig:p_SNR}, since the transmitter is forced to transmit more artificial noise and a stronger energy signal for delivering  energy to the idle receivers. On the other hand,  the normalized maximum channel estimation errors $\sigma_{\mathrm{est}_k}^2$ do not have a large impact on the average secrecy capacity of the system and all schemes are able to guarantee the secrecy QoS because of the proposed robust optimization.

\subsection{Average Total Transmit Power versus Number of Transmit Antennas}
\begin{figure}[t]\hspace*{-4mm}
\subfigure[Average total transmit power.]{\centering
\includegraphics[scale=0.5]{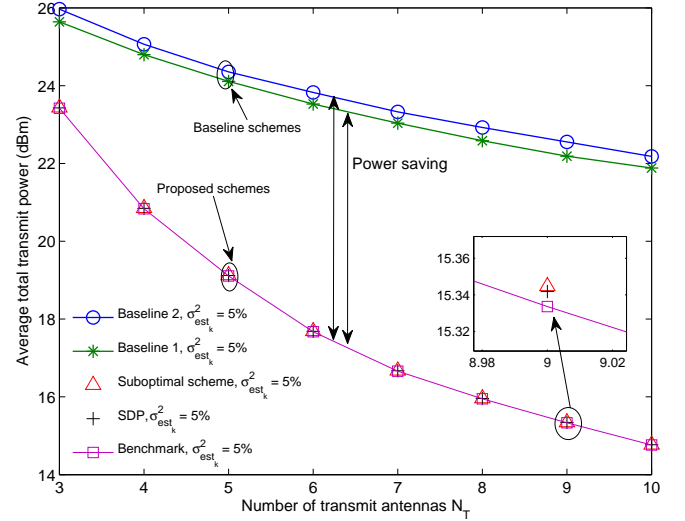}
\label{fig:pt_nt1} }  \subfigure[Average transmit power allocation.]{\centering \hspace*{-4mm}
\includegraphics[scale=0.5]{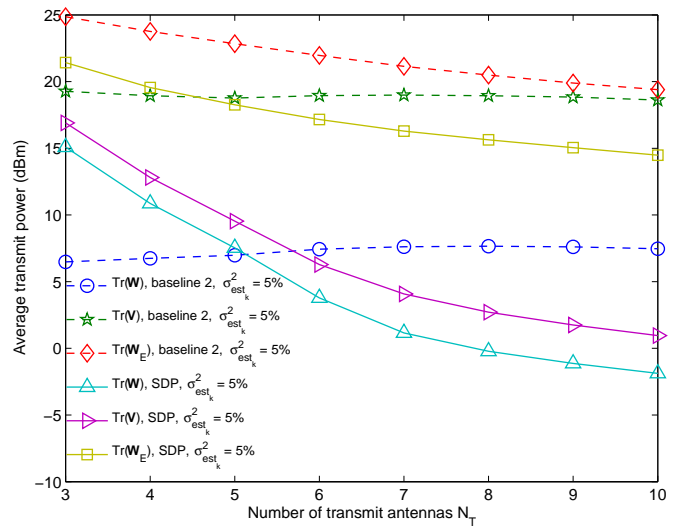}
\label{fig:pt_nt2}}\caption[Optional caption for list of figures]
{Average total transmit power (dBm) and average transmit power allocation (dBm) versus the number of transmit antennas, $N_\mathrm{T}$, for $\Gamma_{\mathrm{req}}=15$ dB, a normalized maximum channel estimation error of $\sigma_{\mathrm{est}_k}^2=5\%$, and different resource allocation schemes. The double-sided arrows indicate the power savings achieved by the proposed schemes compared to the baseline schemes.}\label{fig:pt_nt}
\end{figure}

Figure \ref{fig:pt_nt1} shows the average total transmit power versus the
number of transmit antennas, $N_{\mathrm{T}}$, for $K=4$ legitimate receivers,  different resource allocation schemes, and a normalized maximum channel estimation error of $\sigma_{\mathrm{est}_k}^2=5\%,\forall k$. The minimum required SINR of the desired receiver is set to $\Gamma_{\mathrm{req}}=15$ dB. It is expected that the total transmit power decreases with an increasing number of antennas since extra degrees of freedom can be exploited for resource allocation when  more antennas are available at the transmitter. On the other hand,   the proposed schemes  provide  substantial power savings compared to the  two  baseline schemes due to the optimization of both $\mathbf{W}$ and $\mathbf{W}_{\mathrm{E}}$.  Figure \ref{fig:pt_nt2} depicts the average transmit power allocation to the three components of the transmitted signal, i.e., $\Tr(\mathbf{W}),\Tr(\mathbf{V})$, and $\Tr(\mathbf{W}_{\mathrm{E}})$,  versus the
number of transmit antennas for the proposed SDP resource allocation scheme and baseline scheme 2 under the same system setting as considered in Figure \ref{fig:pt_nt1}.  It can be observed that the amounts of power allocated  to the
information signal and the artificial noise decrease  rapidly for the SDP resource allocation scheme as the number of transmit antennas increases. In fact, the degrees of freedom for resource allocation   increase with the number of transmit antennas. Specifically, with more antennas, the transmitter is able to perform more power efficient  beamforming of the information signal and more effective jamming  of the potential and passive eavesdroppers. Besides, the power allocated to the energy signal decreases with increasing $N_{\mathrm{T}}$ but not as fast as the artificial noise power. This is because the energy signal has less influence on the system performance than the artificial noise. In particular,  the energy signal does not degrade the received SINRs at the potential eavesdroppers. On the other hand,  the powers allocated to the information signal and the artificial noise do not vary significantly with $N_{\mathrm{T}}$ in baseline scheme 2 since the transmitter is unable to fully exploit the extra degrees of freedom introduced by additional transmit antennas.

\begin{figure}[t]\hspace*{-8mm}
\subfigure[Average total transmit power.]{\centering
\includegraphics[scale=0.5]{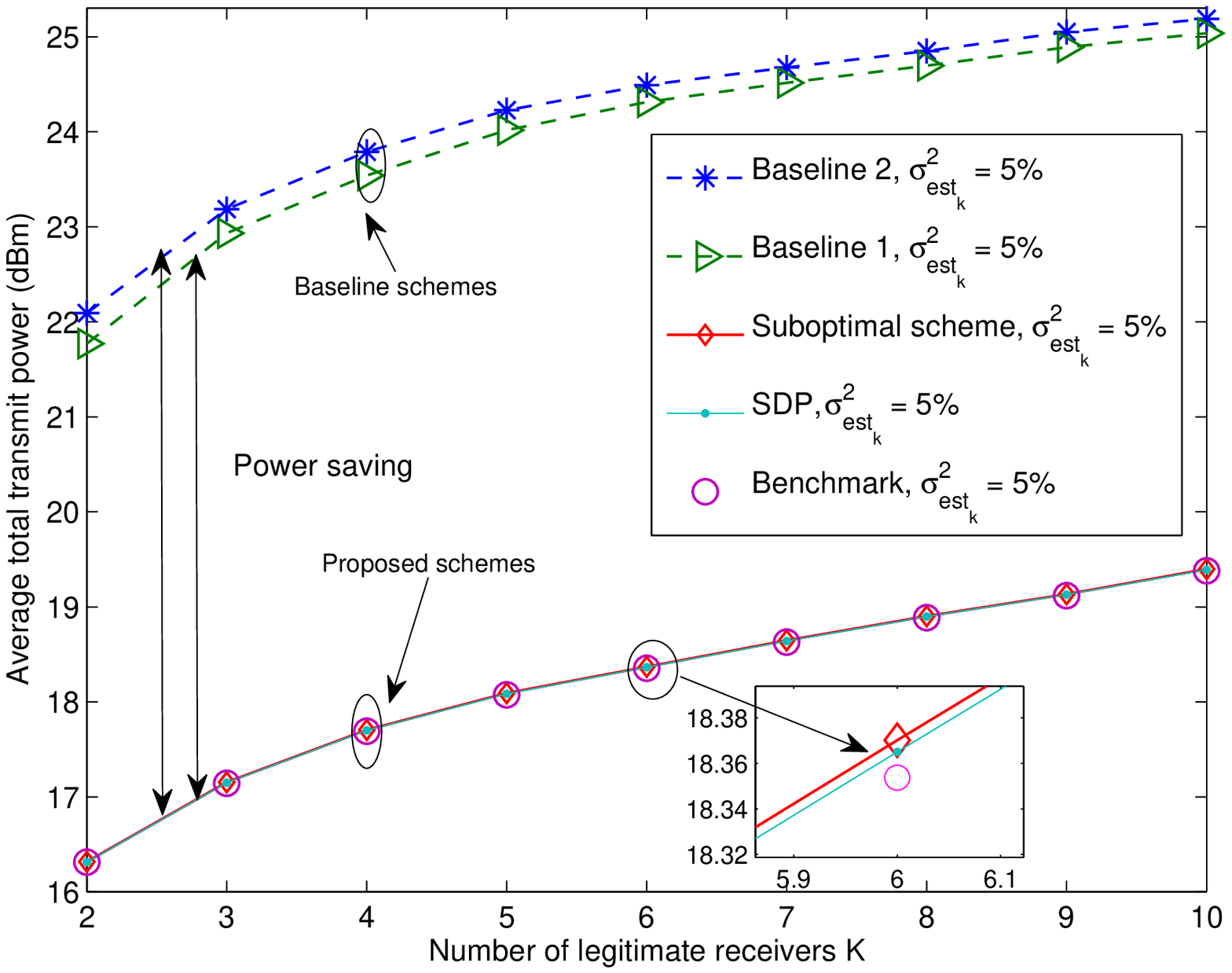}
\label{fig:pt_users1}}\newline \subfigure[Average transmit power allocation.]{\centering \hspace*{-8mm}
\includegraphics[scale=0.5]{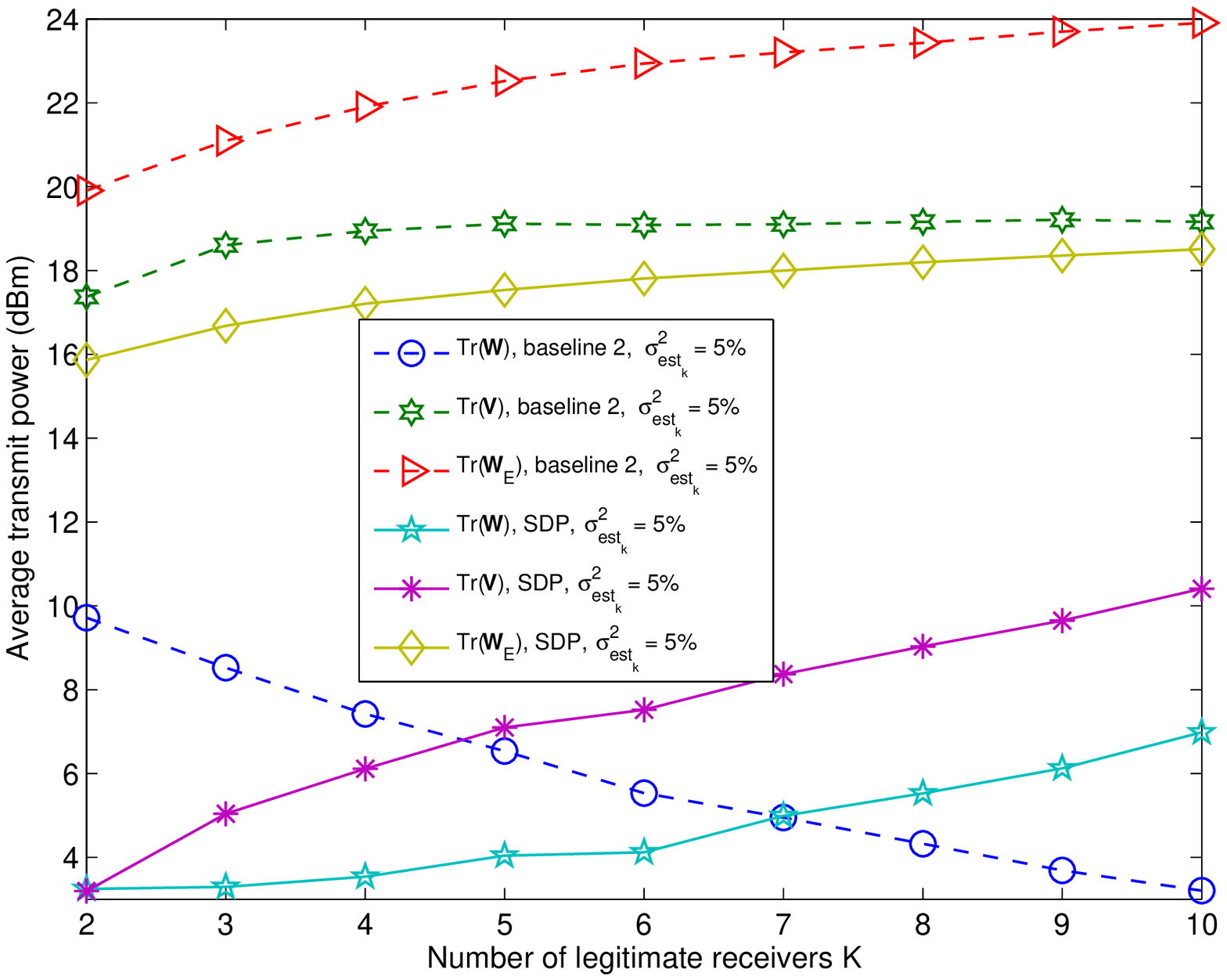}
\label{fig:pt_users2}}\caption[Optional caption for list of figures]
{Average total transmit power (dBm) versus the number of legitimate receivers, $K$, for $\Gamma_{\mathrm{req}}=15$ dB, a normalized maximum channel estimation error of $\sigma_{\mathrm{est}_k}^2=5\%$, and different resource allocation schemes.  The double-sided arrows indicate the power savings achieved by the proposed schemes compared to the baseline schemes.}\label{fig:pt_users}
\end{figure}

\subsection{Average Total Transmit Power versus Number of Legitimate Receivers}
Figure \ref{fig:pt_users1} shows the average total transmit power versus the
number of legitimate receivers, $K$, for  a normalized maximum channel estimation error of $\sigma_{\mathrm{est}_k}^2=5\%,\forall k$, and different resource allocation schemes. The minimum required SINR of the desired receiver is set to $\Gamma_{\mathrm{req}}=15$ dB and there are $N_{\mathrm{T}}=6$ transmit antennas. It is expected that the total transmit power increases with the number of legitimate receivers. The
reason behind this is twofold. First, as the number of legitimate receivers in the system increases, there are more idle receivers requiring power transfer from the transmitter even though some of them are experiencing bad channel conditions. Second, there are more potential eavesdroppers present in the system and thus the transmitter has to generate a higher amount of  artificial noise to guarantee communication secrecy. On the other hand,   the proposed schemes  provide  substantial power savings compared to the  two baseline schemes. Figure \ref{fig:pt_users2} shows the average transmit power allocated the different components of the transmitted signal. We focus on the power allocation   for the proposed SDP resource allocation scheme and baseline scheme 2. It can be observed that  a large portion of power is allocated to the energy signal and artificial noise in both schemes. This is due to the fact that  when there are more potential eavesdroppers (idle legitimate receivers) in the  system, the secrecy QoS and energy harvesting requirements  dominate the system performance as constraints C2 and C5 become more stringent. This translates to a higher demand for artificial noise and energy signal generation (/radiation)  for degrading the channels of the potential eavesdroppers and facilitating efficient power transfer. Besides, it is interesting to note that the amount of power allocated by baseline scheme 2 to the information signal  decreases with  increasing number of legitimate receivers. This result  suggests that with MRT  allocating a larger amount of power to the information signal is not effective in minimizing the total transmit power. Instead,  a larger portion of the transmit  power should be allocated   to the artificial noise and the energy signal for jamming the channel of the potential eavesdroppers   when there are more legitimate receivers in the system.
\begin{figure}[t]\hspace*{-6mm}
\subfigure[Average  total harvested power versus $\Gamma_{\mathrm{req}}$. ]{\centering
\includegraphics[scale=0.5]{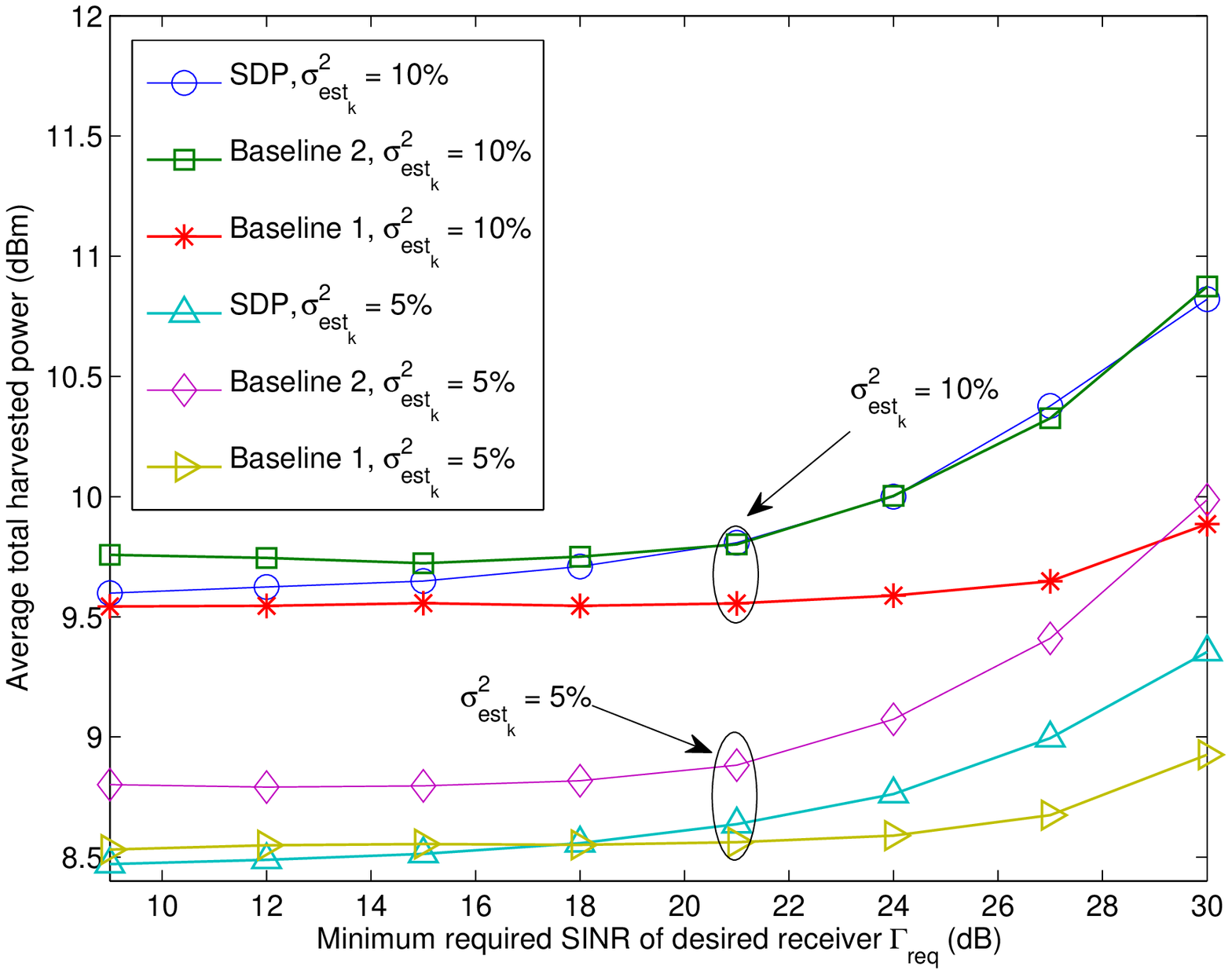}
\label{fig:harvested_PT1} }\newline\subfigure[Average  total harvested power versus $N_{\mathrm{T}}$.]{\centering \hspace*{-6mm}
\includegraphics[scale=0.5]{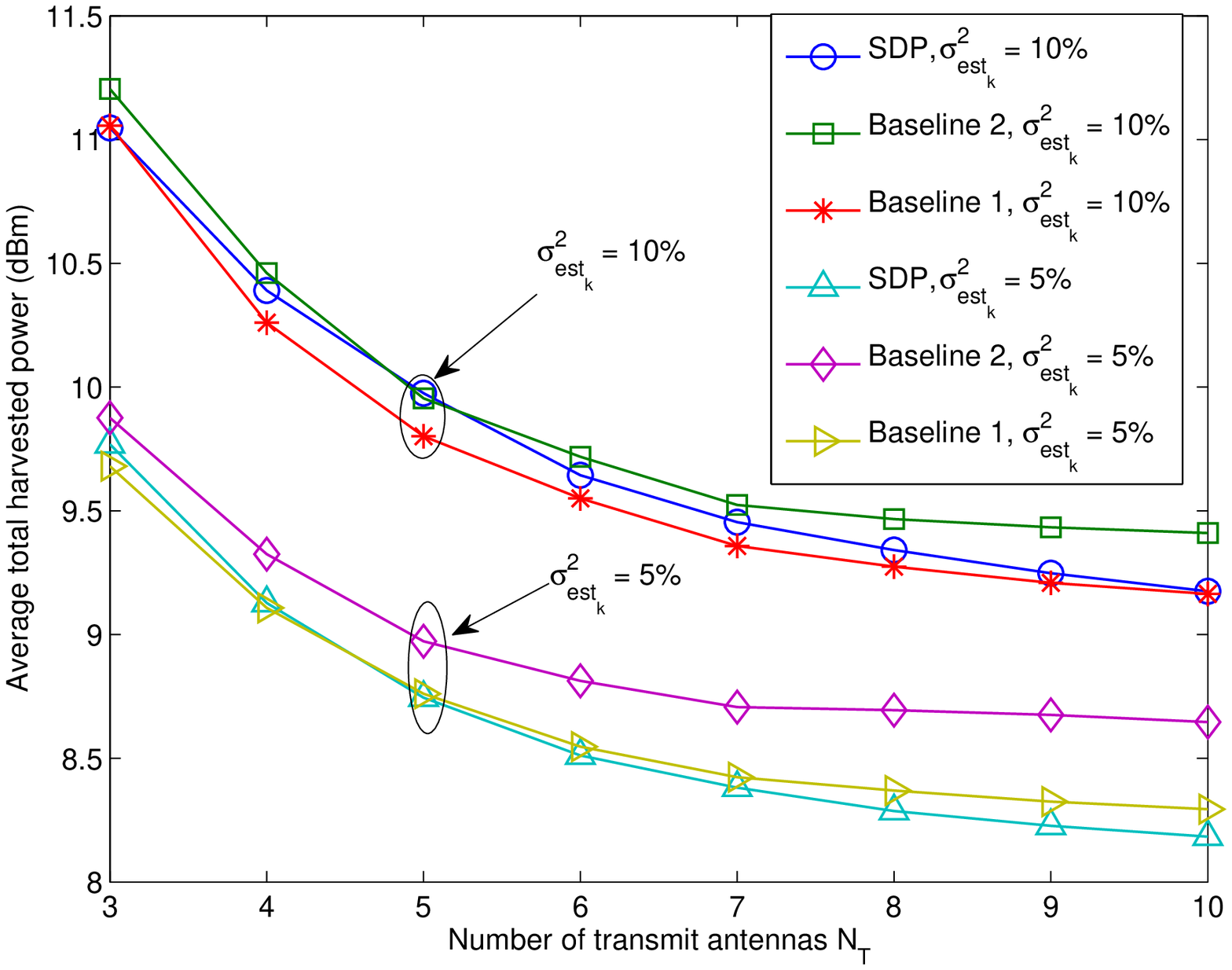}
\label{fig:harvested_PT2} }\caption[Optional caption for list of figures]
{Average  total harvested power (dBm) versus  the minimum required SINR of the desired receiver, $\Gamma_{\mathrm{req}}$ (dB), and   the number of transmit antennas, $N_{\mathrm{T}}$, respectively, for $K=4$ legitimate receivers.}\label{fig:harvested_PT}
\end{figure}

\subsection{Average Total Harvested Power}
Figure \ref{fig:harvested_PT1} shows the average total harvested power versus the minimum required SINR of the desired receiver, $\Gamma_{\mathrm{req}}$, for $K=4$ legitimate receivers, $N_{\mathrm{T}}=6$ transmit antennas,  different normalized maximum channel estimation errors, $\sigma_{\mathrm{est}_k}^2$, and different resource allocation schemes. The average total harvested power is computed by assuming the potential eavesdroppers do not eavesdrop.
 It can be observed that the total average harvested power increases with $\Gamma_{\mathrm{req}}$. On the one hand, the transmitter has to allocate more power to the information bearing signal to satisfy a more stringent requirement on $\Gamma_{\mathrm{req}}$. On the other hand,  the power of the  artificial noise and  energy signal may also increase to reduce the  received SINRs at  both the potential eavesdroppers and the passive eavesdroppers. As a result, for larger $\Gamma_{\mathrm{req}}$, more power is available in the RF and can be harvested by the legitimate receivers. Besides, it can be observed that the average total harvested power increases with increasing $\sigma_{\mathrm{est}_k}^2$. In fact,  to fulfill  the QoS requirements on the minimum power transfer and communication secrecy, a higher amount of transmit power is required  for larger $\sigma_{\mathrm{est}_k}^2$ which leads to a higher energy level  in the RF for energy harvesting.   Figure \ref{fig:harvested_PT2} illustrates the average total harvested power versus the number of transmit antennas, $N_{\mathrm{T}}$, for $K=4$ legitimate receivers, $\Gamma_{\mathrm{req}}=15$ dB, different normalized maximum channel estimation errors, $\sigma_{\mathrm{est}_k}^2$, and different resource allocation schemes.  It can be observed that the average  total harvested power in the system decreases with increasing number of antennas for all considered scenarios. These results suggest that  a lower amount of energy is available in the RF for energy harvesting when the number of transmit antennas increases.  This is due to the fact that with more transmit antennas the direction of beamforming matrix $\mathbf{W}$ can be more accurately steered towards the desired  receiver which reduces the power leakage of $\mathbf{W}$ to the idle receivers for energy harvesting. Besides, by exploiting the extra degrees of freedoms introduced by additional transmit antennas, the artificial noise and energy signal powers  radiated from the transmitter to fulfill the secrecy QoS requirements can be reduced which further decreases the amount of energy available in the RF for energy harvesting. Yet, because of the proposed optimization, the proposed schemes are able to guarantee the minimum required power transfer to the legitimate receivers  for all the cases studied.

\section{Conclusions}\label{sect:conclusion}
In this paper,  we formulated the resource allocation
algorithm design for secure MISO communication  systems with RF energy harvesting receivers as a non-convex optimization problem. The proposed problem formulation advocates the dual use of artificial noise and energy signals  for facilitating   simultaneous secure communication and efficient energy transfer in the presence of both passive eavesdroppers and potential eavesdroppers. Due to the  intractability of the resulting power minimization problem,  the  problem was reformulated by  replacing a non-convex probabilistic constraint with a convex deterministic constraint. Subsequently, an efficient SDP based resource allocation algorithm was proposed to obtain the global optimal solution for the reformulated problem.  Besides a suboptimal low computational complexity resource  allocation scheme was  provided. Simulation results showed
the excellent performance of all the proposed resource allocation schemes. Furthermore, our results also unveiled the power savings enabled by the optimization of  the artificial noise and energy signal generation. An interesting topic for future work is the consideration of multiple antennas eavesdroppers and multiple desired  receivers.
\section{Appendix}
\subsection{Proof of Lemma \ref{lemma:chance_constraint}}\label{sect:proof-lemma1}
Since the channels of the $J$ passive eavesdroppers are independent, the left hand side of constraint C3 in (\ref{eqn:SDP}) can be written as
\begin{eqnarray}\label{eqn:C3-equivalent0}
&&\Pr\Big(\max_{j\in\{1,\ldots,J\}} \Big\{\Gamma_{\mathrm{PE},j}\Big\} \le \Gamma_{\mathrm{tol}}\Big)\notag\\
=&&\prod_{j=1}^{J}\Pr\Big(\Gamma_{\mathrm{PE},j}\le \Gamma_{\mathrm{tol}}\Big).
\end{eqnarray}{
Thus, after some mathematical manipulations, constraint C3 is equivalent to
\begin{eqnarray}\label{eqn:C3-equivalent}
\mbox{C3: }&&\hspace*{-6mm}\Pr\Big(\max_{j\in\{1,\ldots,J\}} \Big\{\Gamma_{\mathrm{PE},j}\Big\} \le \Gamma_{\mathrm{tol}}\Big)\ge \kappa\\
\Leftrightarrow&&\hspace*{-6mm}\Pr\Big(\Gamma_{\mathrm{tol}}\tilde\sigma^2\ge\Tr\Big(\mathbf{\tilde L}(\mathbf{W}-\Gamma_{\mathrm{tol}}\mathbf{W}_\mathrm{E}-\Gamma_{\mathrm{tol}}\mathbf{V})\Big) \Big)\ge \kappa^{1/J}.\notag
\end{eqnarray}}Since the equivalent channels of the passive eavesdroppers are modeled as i.i.d. random variables, we have dropped the index of the passive  eavesdropper channels in (\ref{eqn:C3-equivalent})  without loss of generality.  Now, we focus on the calculation of the probability
$
\Pr\Big(\Gamma_{\mathrm{tol}}\tilde\sigma^2\ge\Tr\Big(\mathbf{\tilde{L}}(\mathbf{W}-\Gamma_{\mathrm{tol}}\mathbf{W}_\mathrm{E}-\Gamma_{\mathrm{tol}}\mathbf{V})\Big) \Big)$. It can be shown that the distribution  of random variable $\Tr\Big(\mathbf{\tilde{L}}(\mathbf{W}-\Gamma_{\mathrm{tol}}\mathbf{W}_\mathrm{E}-\Gamma_{\mathrm{tol}}\mathbf{V})\Big)$ is the same as that of random variable
\begin{eqnarray}\label{eqn:coupled_varibles}
\sum_{n=1}^{N_{\mathrm{T}}} \chi_n^2 \lambda_{n}\big(\mathbf{W}-\Gamma_{\mathrm{tol}}\mathbf{W}_\mathrm{E}-\Gamma_{\mathrm{tol}}\mathbf{V}\big)
\end{eqnarray}
where $\chi_n^2,n\in\{1,\ldots,N_{\mathrm{T}}\},$ are i.i.d. exponential random variables  with probability density function $f_{\chi_n^2}(x) =\exp(-x)$ for $x\ge0$. It can be seen in (\ref{eqn:coupled_varibles}) that the $\chi_n^2$ are coupled with the optimization variable matrices and the distribution of the random variable in (\ref{eqn:coupled_varibles}) depends on the value of the optimization variables $\{\mathbf{W}, \mathbf{W}_\mathrm{E}, \mathbf{V}\}$.  On the other hand, the calculation of the optimization variables depends on the distribution of random variable in (\ref{eqn:coupled_varibles}). This mutual dependence does not facilitate an efficient resource allocation algorithm design. {
 Besides, matrix $\mathbf{W}-\Gamma_{\mathrm{tol}}\mathbf{W}_\mathrm{E}-\Gamma_{\mathrm{tol}}\mathbf{V}$ is in general indefinite and the resulting probability distribution of $\Tr\Big(\mathbf{\tilde{L}}(\mathbf{W}-\Gamma_{\mathrm{tol}}\mathbf{W}_\mathrm{E}-\Gamma_{\mathrm{tol}}\mathbf{V})\Big)$  may not be a convex function. } As a comprise solution, we focus on a smaller convex feasible solution set which facilitates  a tractable solution. We first introduce a trace inequality for the product of two arbitrary Hermitian matrices \cite{JR:trace_inequality}. In particular, for any two Hermitian matrices $\mathbf{A}\in\mathbb{H}^N$ and $\mathbf{B}\in\mathbb{H}^N$, the following inequality holds:
\begin{eqnarray}\label{eqn:inequality}
\sum_{n=1}^N \lambda_n(\mathbf{A})\lambda_{N-n+1}(\mathbf{B})\le \Tr(\mathbf{A}\mathbf{B})\le \sum_{n=1}^N \lambda_n(\mathbf{A})\lambda_n(\mathbf{B}).
\end{eqnarray}

Then, an upper bound for function $\Tr\Big(\mathbf{\tilde{L}}(\mathbf{W}-\Gamma_{\mathrm{tol}}\mathbf{W}_\mathrm{E}-\Gamma_{\mathrm{tol}}\mathbf{V})\Big)$ is given by
\begin{eqnarray}\notag \label{eqn:apply-trace-inequality}
&&\Tr\Big(\mathbf{\tilde{L}}(\mathbf{W}-\Gamma_{\mathrm{tol}}\mathbf{W}_\mathrm{E}-\Gamma_{\mathrm{tol}}\mathbf{V})\Big)\\
&\stackrel{(a)}{\le}&\notag \sum_{n=1}^{N_{\mathrm{T}}} \lambda_n(\mathbf{\tilde{L}})\,\lambda_n(\mathbf{W}-\Gamma_{\mathrm{tol}}\mathbf{W}_\mathrm{E}-\Gamma_{\mathrm{tol}}\mathbf{V})\\
&\notag\stackrel{(b)}{=}&\lambda_{\max}(\mathbf{\tilde{L}})\,\lambda_{\max}(\mathbf{W}-\Gamma_{\mathrm{tol}}\mathbf{W}_\mathrm{E}-\Gamma_{\mathrm{tol}}\mathbf{V})\\
&\stackrel{(c)}{=}&\Tr(\mathbf{\tilde{L}})\,
\lambda_{\max}(\mathbf{W}-\Gamma_{\mathrm{tol}}\mathbf{W}_\mathrm{E}-\Gamma_{\mathrm{tol}}\mathbf{V}),
\end{eqnarray}
where $(a)$ is due to the right hand side inequality in  (\ref{eqn:inequality}) while $(b)$ and $(c)$ are due to the fact that $\mathbf{\tilde{L}}$ is a rank-one positive semi-definite matrix. Thus, by combining (\ref{eqn:C3-equivalent0}), (\ref{eqn:C3-equivalent}), and (\ref{eqn:apply-trace-inequality}), we obtain the following inequality:
\begin{eqnarray}
\hspace*{-6mm}&&\hspace*{-6mm}\Pr\Big(\Tr\Big(\mathbf{\tilde L}(\mathbf{W}-\Gamma_{\mathrm{tol}}\mathbf{W}_\mathrm{E}-\Gamma_{\mathrm{tol}}\mathbf{V})\Big)\le \Gamma_{\mathrm{tol}}\tilde\sigma^2 \Big)\notag\\
\hspace*{-6mm}\ge&& \hspace*{-6mm} \Pr\Big(\lambda_{\max}(\mathbf{W}-\Gamma_{\mathrm{tol}}\mathbf{W}_\mathrm{E}-\Gamma_{\mathrm{tol}}\mathbf{V})\Tr(\mathbf{\tilde{L}})\le \Gamma_{\mathrm{tol}}\tilde\sigma^2
\Big) . \end{eqnarray} {
As a result, by setting $\Pr\Big(\lambda_{\max}(\mathbf{W}-\Gamma_{\mathrm{tol}}\mathbf{W}_\mathrm{E}-\Gamma_{\mathrm{tol}}\mathbf{V})\Tr(\mathbf{\tilde{L}})\le \Gamma_{\mathrm{tol}}\tilde\sigma^2
\Big)\ge \kappa^{1/J}$, we have
 \begin{subequations}\label{eqn:implication_0}
 \begin{eqnarray}\label{eqn:implication}
&&\hspace*{-6mm}\Pr\Big(\lambda_{\max}(\mathbf{Q})\Tr(\mathbf{\tilde{L}})\le \Gamma_{\mathrm{tol}}\tilde\sigma^2
\Big)\ge \kappa^{1/J}\\
\hspace*{-6mm}\stackrel{(d)}{\Longleftrightarrow}&&\hspace*{-6mm} \Pr\Big(\frac{\lambda_{\max}\big(\mathbf{Q}\big)}{\Gamma_{\mathrm{tol}}\tilde\sigma^2}\le \frac{1}{\Tr(\mathbf{\tilde{L}}) }\Big)\ge \kappa^{1/J} \\
\hspace*{-6mm}\stackrel{(e)}{\Longleftrightarrow}&&\hspace*{-6mm} \Pr\Big(\frac{\lambda_{\max}\big(\mathbf{Q}\big)}{\Gamma_{\mathrm{tol}}\tilde\sigma^2}\ge \frac{1}{\Tr(\mathbf{\tilde{L}}) }\Big)\le 1-\kappa^{1/J} \\ \label{eqn:implication_eq1}
\hspace*{-6mm}\Longleftrightarrow&&\hspace*{-6mm} \Phi^{-1}_{N_{\mathrm{T}}}(1-\kappa^{1/J})\Gamma_{\mathrm{tol}}\tilde\sigma^2\ge \lambda_{\max}\big(\mathbf{Q}\big)\label{eqn:implication2}\\ \label{eqn:implication_eq2}
\hspace*{-6mm}\Longleftrightarrow&&\hspace*{-6mm}
\mathbf{I}_{\mathrm{N_T}}\Big(\Phi^{-1}_{N_{\mathrm{T}}}(1-\kappa^{1/J})\Gamma_{\mathrm{tol}}\tilde\sigma^2\Big)\succeq \mathbf{Q}\\
\hspace*{-6mm}\Longrightarrow&&\hspace*{-6mm}\Pr\Big(\Gamma_{\mathrm{tol}}\tilde\sigma^2\ge\Tr\Big(\mathbf{\tilde L}\mathbf{Q}\Big) \Big)\ge \kappa^{1/J}\\
\hspace*{-6mm}\Longleftrightarrow&&\hspace*{-6mm}\Pr\Big(\max_{j\in\{1,\ldots,J\}} \Big\{\Gamma_{\mathrm{PE},j}\Big\} \le \Gamma_{\mathrm{tol}}\Big)\hspace*{-0.5mm}\ge\hspace*{-0.5mm} \kappa,
\end{eqnarray}
 \end{subequations}
where $\mathbf{Q}=\mathbf{W}-\Gamma_{\mathrm{tol}}\mathbf{W}_\mathrm{E}-\Gamma_{\mathrm{tol}}\mathbf{V}$.  $(d)$ and $(e)$ are due to  the positive definiteness of matrix $\mathbf{\tilde{L}}$ and a basic property of probability, respectively. $\Phi^{-1}_{N_{\mathrm{T}}}(\cdot)$ in (\ref{eqn:implication2}) denotes the inverse
cumulative distribution function (c.d.f.) of an inverse central
chi-square random variable with 2$N_\mathrm{T}$ degrees of freedom. We note that the deviations in (\ref{eqn:implication_0}) do not require matrix $\mathbf{W}-\Gamma_{\mathrm{tol}}\mathbf{W}_\mathrm{E}-\Gamma_{\mathrm{tol}}\mathbf{V}$ to be a positive semidefinite matrix.  Besides, the non-smoothness of $\lambda_{\max}(\cdot)$ is overcome by replacing (\ref{eqn:implication_eq1}) with its equivalent expression in (\ref{eqn:implication_eq2}).} Furthermore, random variable $\mathbf{\tilde L}$ is decoupled from the  optimization variables, cf.  (\ref{eqn:apply-trace-inequality}). Thus,  the implication in (\ref{eqn:implication2}) is applicable to any continuous channel distribution by replacing $\Phi^{-1}_{N_{\mathrm{T}}}(\cdot)$ with an inverse c.d.f. with respect to  the corresponding distribution. In practice,  the inverse function of the
inverse central chi-square c.d.f. can be evaluated directly or be
stored in a lookup table in a practical implementation.

\subsection{Proof of Theorem 1}{
The proof of Theorem 1 utilizes the results from  \cite[Proposition 4.1]{JR:rui_zhang}. In the first part, we  show the structure of the optimal solution $\mathbf{W}^*$. Then, in the second part, we propose a method for constructing a solution $\{\mathbf{\widetilde W^*},\mathbf{\widetilde V^*,}  \mathbf{\widetilde W}_\mathrm{E}^*, \widetilde \rho^*,\boldsymbol{\widetilde \delta}^*,  \boldsymbol{\widetilde \nu}^*\}$ that achieves the same objective value as $\{\mathbf{ W^*},\mathbf{ V^*,}  \mathbf{ W}_\mathrm{E}^*,  \rho^*,\boldsymbol{ \delta}^*,  \boldsymbol{ \nu}^*\}$ and admits a rank-one $ \mathbf{\widetilde W}^*$. It can be shown that the relaxed version of problem (\ref{eqn:SDP-robust}) is jointly convex with respect to the optimization variables and satisfies Slater's constraint qualification. As a result, the  Karush-Kuhn-Tucker (KKT) conditions  are necessary and sufficient conditions \cite{book:convex} for the solution of the relaxed problem. {
The Lagrangian function  of  (\ref{eqn:SDP-robust})  is given by
\begin{eqnarray}\label{eqn:Lagrangian}\notag{\cal
L}\hspace*{-1mm}&=&\hspace*{-3mm} \Tr(\mathbf{W})-\Tr(\mathbf{WY}) - \sum_{k=1}^{K-1}\Tr\Big(\mathbf{S}_{\mathrm{C}_{2_k}}\big(\mathbf{W},\mathbf{V},\delta_k\big)\mathbf{D}_{\mathrm{C}_{2_k}}\Big)\notag\\
\notag\hspace*{-1mm}&-&\hspace*{-3mm}
\sum_{k=1}^{K-1}\Tr\Big(\mathbf{S}_{\mathrm{C}_{5_k}}\big(\mathbf{W},\mathbf{V},\mathbf{W}_\mathrm{E},\nu_k\big)\mathbf{D}_{\mathrm{C}_{5_k}}\Big)\\
\hspace*{-1mm}&+&\hspace*{-3mm}  \mu\Big(\frac{ P_{\min}}{\eta(1-\rho)}-\Tr(\mathbf{H}\mathbf{W})\Big)+ \beta\Big(\frac{\Gamma_{\mathrm{req}}\sigma_{\mathrm{s}}^2 }{\rho}\hspace*{-0.5mm}-\hspace*{-0.5mm}\Tr(\mathbf{H}\mathbf{W})\Big)\notag\\
\hspace*{-1mm}&+&\hspace*{-3mm} \sum_{n=1}^{N_{\mathrm{T}}}\theta_n \Tr\Big(\mathbf{\Psi}_n\mathbf{W}\hspace*{-0.5mm}\Big)\hspace*{-0.5mm}+\hspace*{-0.5mm} \Tr\Big(\mathbf{W}\mathbf{D}_{\overline{\mathrm{C}_{3}}}\Big)+\Delta,
\end{eqnarray}
where $\Delta$ denotes the collection of terms that only involve variables that are not relevant for the proof. In (\ref{eqn:Lagrangian}), $\beta\ge 0$ is the Lagrange multiplier for the minimum required SINR of the desired receiver in C1. $\mathbf{D}_{\mathrm{C}_{2_k}}\succeq \mathbf{0},\forall k\in\{1,\,\ldots,\,K-1\},\mathbf{D}_{\overline{\mathrm{C}_{3}}}\succeq \mathbf{0}$, and $\mathbf{D}_{\mathrm{C}_{5_k}}\succeq \mathbf{0},\forall k\in\{1,\,\ldots,\,K-1\},$ are the Lagrange multiplier matrices corresponding to constraints C2,  $\overline{\mbox{C3}}$, and  C5, respectively. Lagrange multiplier  $\mu\ge 0$  corresponds to the minimum required power transfer to the desired receiver in C4.
 $\boldsymbol{\theta}$ is the Lagrange multiplier vector
corresponding to constraint C6 on the per-antenna maximum transmit power with elements $\theta_n\ge0,\forall n\in\{1,\ldots,N_{\mathrm{T}}\}$. Matrix $\mathbf{Y}\succeq \mathbf{0}$ is the Lagrange multiplier matrix for the positive semi-definite constraint on matrix $\mathbf{W}$  in C8.} Considering (\ref{eqn:Lagrangian}), the KKT conditions for the optimal $\mathbf{W}^*$ are given by:
\begin{eqnarray}\label{eqn:KKT}\mathbf{Y}^*,\mathbf{D}_{\mathrm{C}_{2_k}}^*,\mathbf{D}_{\overline{\mathrm{C}_{3}}}^*,\mathbf{D}_{\mathrm{C}_{5_k}}^*
\hspace*{-3mm}&\succeq&\hspace*{-3mm} \mathbf{0},\quad\mu^*,\,\beta^*,\,\,\theta_n^*\ge 0,\\
 \mathbf{Y^*W^*}\hspace*{-3mm}&=&\hspace*{-3mm}\mathbf{0}, \label{eqn:KKT-complementarity}\\
\nabla_{\mathbf{W}}{\cal L}\hspace*{-3mm}&=&\hspace*{-3mm} \mathbf{0}, \label{eqn:subgradient}\\ \label{eqn:optimal_rho}
\frac{\partial {\cal L}}{\partial \rho^*}=0\Rightarrow\rho^*\hspace*{-3mm}&=&\hspace*{-3mm}\frac{\sqrt{\beta^*\sigma_{\mathrm{s}}^2 \Gamma_{\mathrm{req}}}}{\sqrt{\beta^*\sigma_{\mathrm{s}}^2 \Gamma_{\mathrm{req}}}+\sqrt{\frac{\mu^* P_{\min}}{\eta}}},
\end{eqnarray}
where $\mathbf{Y}^*,\mathbf{D}_{\mathrm{C}_{2_k}}^*,\mathbf{D}_{\overline{\mathrm{C}_{3}}}^*,\mathbf{D}_{\mathrm{C}_{5_k}}^*,\mu^*,\beta^*$ and $\theta_n^*$, are the optimal Lagrange multipliers for the dual problem of (\ref{eqn:SDP-robust}). It can be observed from (\ref{eqn:optimal_rho}) that constraint $\mbox{C7: } 0\le \rho\le 1$ is  automatically satisfied. Besides,  $\mu^*,\beta^*>0$  must holds for $\Gamma_{\mathrm{req}}>0$ and $P^{\min}>0$. On the other hand, (\ref{eqn:KKT-complementarity}) is the complementary slackness condition and is satisfied when the columns of $\mathbf{W}^*$ lie in the null space of $\mathbf{Y}^*$. Thus, the KKT condition in (\ref{eqn:subgradient}) can be expressed as
\begin{eqnarray}\label{eqn:KKT-gradient-equivalent}\notag
&&\hspace*{-5mm}\mathbf{I}_{N_{\mathrm{T}}}+\mathbf{D}_{\overline{\mathrm{C}_{3}}}^*+\sum_{n=1}^{N_{\mathrm{T}}}\theta_n^*\mathbf{\Psi}_n+\sum_{k=1}^{K-1} \mathbf{U}_{\mathbf{g}_k}\Big(\frac{\mathbf{D}^*_{\mathrm{C}_{2_k}}}{\Gamma_{\mathrm{tol}_k}}-\mathbf{D}_{\mathrm{C}_{5_k}}^*\Big)\mathbf{U}_{\mathbf{g}_k}^H
\\
=&&\hspace*{-5mm}\mathbf{Y}^*+(\mu^*+\beta^*)\mathbf{H}.
\end{eqnarray}
For notational simplicity, we define
\begin{eqnarray}\label{eqn:B_matrix}
\mathbf{B}^*&=&\mathbf{I}_{N_{\mathrm{T}}}+\mathbf{D}_{\overline{\mathrm{C}_{3}}}^* +\sum_{n=1}^{N_{\mathrm{T}}}\theta_n^*\mathbf{\Psi}_n\notag\\
&+&\sum_{k=1}^{K-1} \mathbf{U}_{\mathbf{g}_k}\Big(\frac{\mathbf{D}^*_{\mathrm{C}_{2_k}}}{\Gamma_{\mathrm{tol}_k}}-\mathbf{D}^*_{\mathrm{C}_{5_k}}\Big)
\mathbf{U}_{\mathbf{g}_k}^H.
\end{eqnarray}
 From the complementary slackness condition in (\ref{eqn:KKT-complementarity}),  the columns of $\mathbf{W}^*$ have to lie in the null space of $\mathbf{Y}^*$ for $\mathbf{W}^*\ne \mathbf{0}$.  Thus, we study  the rank and null space of $\mathbf{Y}^*$ for obtaining the structure of the optimal solution $\mathbf{W}^*$. From (\ref{eqn:KKT-gradient-equivalent}) and (\ref{eqn:B_matrix}), we obtain for the Lagrange multiplier matrix
\begin{eqnarray}\label{eqn:Y}
\mathbf{Y}^*=\mathbf{B}^*-(\mu^*+\beta^*)\mathbf{H},
\end{eqnarray}
where $(\mu^*+\beta^*)\mathbf{H}$ is a rank-one matrix since $\mu^*,\beta^*>0$. Without loss of generality, we define $r=\Rank(\mathbf{B}^*)$ and the orthonormal basis of the null space of $\mathbf{B}^*$ as $\mathbf{\mathbf{\Upsilon}}\in\mathbb{C}^{N_{\mathrm{T}}\times (N_{\mathrm{T}}-r)}$ such that $\mathbf{B}^*\mathbf{\Upsilon}=\mathbf{0}$ and $\Rank(\mathbf{\Upsilon})=N_{\mathrm{T}}-r$. Let $\boldsymbol{\varrho}_t\in\mathbb{C}^{N_{\mathrm{T}}\times 1}$, $1\le t\le N_{\mathrm{T}}-r$, denote the $t$-th column of $\mathbf{\Upsilon}$. By exploiting \cite[Proposition 4.1]{JR:rui_zhang}, it can be shown that $\mathbf{H}\mathbf{\Upsilon}=\mathbf{0}$ and  we can express the optimal solution of  $\mathbf{W}^*$  as
\begin{eqnarray}\label{eqn:general_structure}
\mathbf{W}^*=\sum_{t=1}^{N_{\mathrm{T}}-r} \gamma_t \boldsymbol{\varrho}_t \boldsymbol{\varrho}_t^H  + f\mathbf{u}\mathbf{u}^H,
\end{eqnarray}
where $ \gamma_t \ge 0,\forall t, f>0,f\in\mathbb{R}$, are positive scaling constants  and $\mathbf{u}\in \mathbb{C}^{N_{\mathrm{T}}\times 1}$, $\norm{\mathbf{u}}=1$, satisfies $\mathbf{u}^H\mathbf{\Upsilon}=\mathbf{0}$.
}
\newcounter{mytempeqncnt}
\begin{figure*}[ht]\setcounter{mytempeqncnt}{\value{equation}}
\setcounter{equation}{44}
\begin{eqnarray}\label{eqn:equivalent_objective}
 \hspace*{-5mm}\mbox{Objective value:} &&\hspace*{-5mm}\Tr(\mathbf{\widetilde W^*})+\Tr(\mathbf{\widetilde V^*})+\Tr( \mathbf{\widetilde W^*}_\mathrm{E})= \Tr(\mathbf{W^*})+\Tr(\mathbf{V^*})+\Tr(\mathbf{ W}_\mathrm{E}^*)\\
 \label{eqn:C1_new_solution}
\mbox{C1:}&& \hspace*{-5mm} \frac{\Tr(\mathbf{\widetilde W^*}\mathbf{H})}{\sigma_{\mathrm{ant}}^2+\Tr(\mathbf{H}\mathbf{\widetilde V^*})+\frac{\sigma_{\mathrm{s}}^2}{\rho}}= \frac{\Tr((\mathbf{W}^*-\sum_{t=1}^{N_{\mathrm{T}}-r} \gamma_t \boldsymbol{\varrho}_t \boldsymbol{\varrho}_t^H)\mathbf{H})}{\sigma_{\mathrm{ant}}^2+\Tr(\mathbf{H}(\mathbf{ V^*}+\sum_{t=1}^{N_{\mathrm{T}}-r} \gamma_t \boldsymbol{\varrho}_t \boldsymbol{\varrho}_t^H))+\frac{\sigma_{\mathrm{s}}^2}{\rho}}\notag\\
  =&& \hspace*{-5mm}\frac{\Tr(\mathbf{W}^*\mathbf{H})}{\sigma_{\mathrm{ant}}^2+\Tr(\mathbf{H}\mathbf{ V^*})+\frac{\sigma_{\mathrm{s}}^2}{\rho}}\ge  \Gamma_{\mathrm{req}},\\
  \mbox{C2:}&&\hspace*{-5mm} \mathbf{S}_{\mathrm{C}_{2_k}}\hspace*{-0.5mm}(\mathbf{\widetilde W^*},\mathbf{\widetilde V^*},\widetilde\delta_k\hspace*{-0.5mm})\succeq\mathbf{S}_{\mathrm{C}_{2_k}}\hspace*{-0.5mm}(\mathbf{ W^*},\mathbf{ V^*},\delta_k\hspace*{-0.5mm})\hspace*{-0.5mm}+\hspace*{-0.5mm}\mathbf{U}_{\mathbf{g}_k}^H\Big[\hspace*{-0.5mm}\sum_{t=1}^{N_{\mathrm{T}}-r} \gamma_t \boldsymbol{\varrho}_t \boldsymbol{\varrho}_t^H \hspace*{-0.5mm}\Big] \mathbf{U}_{\mathbf{g}_k}(1\hspace*{-0.5mm}+\hspace*{-0.5mm}\frac{1}{\Gamma_{\mathrm{tol}_k}}) \succeq \mathbf{0},\forall k,\\ \label{eqn:C3:lambda_max}
  \overline{\mbox{C3}}\mbox{:}&&\hspace*{-5mm} { \lambda_{\max}\big(\mathbf{\widetilde W^*}\hspace*{-0.5mm}-\hspace*{-0.5mm}\Gamma_{\mathrm{tol}} \mathbf{\widetilde W^*}_\mathrm{E}
\hspace*{-0.5mm}-\hspace*{-0.5mm}\Gamma_{\mathrm{tol}}\mathbf{\widetilde V^*}\big)\le \lambda_{\max}\big(\mathbf{ W^*}\hspace*{-0.5mm}-\hspace*{-0.5mm}\Gamma_{\mathrm{tol}}\mathbf{W}_\mathrm{E}^*
\hspace*{-0.5mm}-\hspace*{-0.5mm}\Gamma_{\mathrm{tol}}\mathbf{V}^*\big)\le \Gamma_{\mathrm{tol}}\tilde\sigma^2\Phi^{-1}_{N_{\mathrm{T}}}(1-\kappa^{1/J})},\\
 \mbox{C4:}\notag&&\hspace*{-5mm}\Tr(\mathbf{H}\mathbf{\widetilde W^*})+\Tr(\mathbf{H}\mathbf{\widetilde W^*}_\mathrm{E})+\Tr(\mathbf{H}\mathbf{\widetilde V^*})+\sigma_{\mathrm{ant}}^2\\
 =&&\hspace*{-5mm}\Tr(\mathbf{H}\mathbf{ W^*})+\Tr(\mathbf{H}\mathbf{ W}_\mathrm{E}^*)+\Tr(\mathbf{H}\mathbf{ V^*})+\sigma_{\mathrm{ant}}^2 \ge\frac{ P_{\min}}{(1-\rho^*)\eta}=\frac{ P_{\min}}{(1-\tilde \rho^*)\eta},\\
  \mbox{C5:}&&\hspace*{-5mm}\mathbf{S}_{\mathrm{C}_{5_k}}(\mathbf{\widetilde W^*},\mathbf{\widetilde V^*},\mathbf{\widetilde W^*}_\mathrm{E},\widetilde \nu_k^*)=\mathbf{S}_{\mathrm{C}_{5_k}}(\mathbf{W^*},\mathbf{V^*},\mathbf{ W}_\mathrm{E}^*,\nu_k^*)\succeq \mathbf{0},\forall k,\\
    \mbox{C6:}&&\hspace*{-5mm} \Tr\Big(\mathbf{\Psi}_n\big(\mathbf{W}^*+\mathbf{V}^*+\mathbf{ W}_\mathrm{E}^*\big)\Big)
    =\Tr\Big(\mathbf{\Psi}_n\big(\mathbf{\widetilde  W^*}+\mathbf{\widetilde V^*}+\mathbf{\widetilde W^*}_{\mathrm{E}}\big)\Big)\le P_{\max_n}, \forall n,\\
    \mbox{C7:}&&\hspace*{-5mm}\,\, 0\le\widetilde\rho^*\le 1, \,\,\,\mbox{C8:}\,\, \mathbf{\widetilde W ^*}\succeq \mathbf{0}, \mathbf{\widetilde V^*}\succeq \mathbf{0},\mathbf{\widetilde W^*}_{\mathrm{E}}\succeq \mathbf{0},\,\,\,
   \mbox{C10: }\widetilde\delta_k^*,\widetilde\nu_k^* \ge 0,\forall k.  \label{eqn:C10_new_solution}
\end{eqnarray} \hrulefill \vspace*{-4mm}  \addtocounter{mytempeqncnt}{8}
\setcounter{equation}{42}
\end{figure*}

In the second part of the proof, for $\Rank(\mathbf{W}^*)>1$,  we construct another solution of the relaxed version of problem (\ref{eqn:SDP-robust}),  $\{\mathbf{\widetilde W^*},\mathbf{\widetilde V^*,}  \mathbf{\widetilde W^*}_\mathrm{E}, \widetilde \rho^*,$ $\boldsymbol{\widetilde \delta}^*,  \boldsymbol{\widetilde \nu}^*\}$, based on  (\ref{eqn:general_structure}).
Let
\begin{eqnarray}\label{eqn:rank-one-structure}\mathbf{\widetilde W^*}&=&f\mathbf{u}\mathbf{u}^H=\mathbf{W}^*-\sum_{t=1}^{N_{\mathrm{T}}-r} \gamma_t \boldsymbol{\varrho}_t \boldsymbol{\varrho}_t^H, \\
 \mathbf{\widetilde V^*}&=&\mathbf{ V^*}+\sum_{t=1}^{N_{\mathrm{T}}-r} \gamma_t \boldsymbol{\varrho}_t \boldsymbol{\varrho}_t^H,\notag\\
\mathbf{\widetilde W}_\mathrm{E}^*&=& \mathbf{ W}_\mathrm{E}^*,\,\,\, \widetilde \rho^*=\rho^*, \,\,\,\boldsymbol{\widetilde \delta}^*=\boldsymbol{\delta}^*,\,\,\, \mbox{and} \,\,\,  \boldsymbol{\widetilde \nu}^*=\boldsymbol{ \nu}^*. \label{eqn:rank-one-structure2}
\end{eqnarray}
Then, we substitute $\{\mathbf{\widetilde W^*},\mathbf{\widetilde V^*,}  \mathbf{\widetilde W}_\mathrm{E}^*, \widetilde \rho^*, \boldsymbol{\widetilde \delta}^*,  \boldsymbol{\widetilde \nu}^*\}$ into the objective function and constraints in (\ref{eqn:SDP-robust}) which yields the  equations (\ref{eqn:equivalent_objective})--(\ref{eqn:C10_new_solution}) on the top of this page, where (\ref{eqn:C3:lambda_max}) is due to (\ref{eqn:implication_eq1}) and (\ref{eqn:implication_eq2}).
It can be seen from (\ref{eqn:equivalent_objective}) that the solution set $\{\mathbf{\widetilde W^*},\mathbf{\widetilde V^*,}  \mathbf{\widetilde W}_\mathrm{E}^*, \widetilde \rho^*, \boldsymbol{\widetilde \delta}^*,  \boldsymbol{\widetilde \nu}^*\}$ achieves the same optimal value as the optimal solution $\{\mathbf{W^*,V^*,}\mathbf{W}_\mathrm{E}^*, \rho^*,\boldsymbol \delta^*, \boldsymbol \nu^*\}$ and satisfies all the constraints, cf. (\ref{eqn:C1_new_solution})--(\ref{eqn:C10_new_solution}). Thus, $\{\mathbf{\widetilde W^*},\mathbf{\widetilde V^*,}  \mathbf{\widetilde W}_\mathrm{E}^*, \widetilde \rho^*, \boldsymbol{\widetilde \delta}^*,  \boldsymbol{\widetilde \nu}^*\}$  is also an optimal solution but with   $\Rank(\mathbf{\widetilde W^*})=1$, cf. (\ref{eqn:rank-one-structure}).    \qed

\begin{Remark}\label{Remark:construction} We note that $\{\mathbf{\widetilde W^*},\mathbf{\widetilde V^*,}  \mathbf{\widetilde W}_\mathrm{E}^*, \widetilde \rho^*, \boldsymbol{\widetilde \delta}^*,  \boldsymbol{\widetilde \nu}^*\}$ can be obtained by substituting  (\ref{eqn:rank-one-structure}) and (\ref{eqn:rank-one-structure2}) into the relaxed version of (\ref{eqn:SDP-robust}). Then, we solve the resulting  convex optimization problem for  $f$ and $\gamma_t$. On the other hand, the constructed solution set with $\Rank(\mathbf{\widetilde W^*})=1$ is not  unique. In (\ref{eqn:rank-one-structure}), we construct $\mathbf{\widetilde V^*}$ by modifying the artificial noise structure of the original solution,  $\mathbf{ V^*}$,   such that the constructed beamforming matrix $\mathbf{\widetilde W^*}$  is  rank-one. In fact, for achieving a rank-one $\mathbf{\widetilde W^*}$, a general  structure for constructing the artificial noise and the energy signal  can be expressed as $\mathbf{\widetilde V^*}=\mathbf{ V^*}+\sum_{t=1}^{N_{\mathrm{T}}-r} \pi_{t}\gamma_t \boldsymbol{\varrho}_t \boldsymbol{\varrho}_t^H$
 and $\mathbf{\widetilde W}_\mathrm{E}^*= \mathbf{ W}_\mathrm{E}^*+\sum_{t=1}^{N_{\mathrm{T}}-r} (1-\pi_{t})\gamma_t \boldsymbol{\varrho}_t \boldsymbol{\varrho}_t^H$, respectively, for any $0\le \pi_{t}\le 1,\forall t\in\{1,\ldots,N_{\mathrm{T}}-r\}$.
Yet, the non-uniqueness of the constructed solution does not play an important role in this paper since all solutions achieve the same optimal objective value while satisfying all the constraints.
\end{Remark}

\subsection{Proof of Proposition 1}
We prove that $\Rank(\mathbf{W}^*)=1$ when $\frac{\mathbf{D}^*_{\mathrm{C}_{2_k}}}{\Gamma_{\mathrm{tol}_k}}-\mathbf{D}^*_{\mathrm{C}_{5_k}}\succeq \mathbf{0},\forall k$ holds.
First, we post-multiply both sides of (\ref{eqn:Y}) by $\mathbf{W}^*$ which yields \setcounter{equation}{52}
\begin{eqnarray}\label{eqn:pre-rank_equality}
\mathbf{B}^*\mathbf{W}^*\hspace*{-1mm}&=&\hspace*{-1mm}\Big(\mathbf{Y}^*+(\mu^*+\beta^*)\mathbf{H}\Big)\mathbf{W}^*\notag\\
\hspace*{-1mm}&\stackrel{(a)}{=}&\hspace*{-1mm}(\mu^*+\beta^*)\mathbf{H} \mathbf{W}^*,
\end{eqnarray}
where $(a)$ is due to the complementary slackness condition in (\ref{eqn:KKT-complementarity}).
Since $\frac{\mathbf{D}^*_{\mathrm{C}_{2_k}}}{\Gamma_{\mathrm{tol}_k}}-\mathbf{D}^*_{\mathrm{C}_{5_k}}\succeq \mathbf{0},\forall k$ holds by assumption,  $\mathbf{B}^*$ is a positive definite matrix. Then, by exploiting (\ref{eqn:pre-rank_equality}) and a basic rank inequality, we have
\begin{eqnarray}\label{eqn:rank_equality}
\Rank(\mathbf{W}^*)\hspace*{-2.5mm}&=&\hspace*{-2.5mm}\Rank(\mathbf{B^*W^*})=\Rank((\mu^*+\beta^*)\mathbf{H}\mathbf{W}^*)\notag\\
\hspace*{-2.5mm}&=&\hspace*{-2.5mm}\min\{\Rank(\mathbf{W}^*),\Rank\big((\mu^*+\beta^*)\mathbf{H}\big)\hspace*{-0.5mm}\}.
\end{eqnarray}
We note that $\Rank(\mathbf{H})=1$ and $\Rank\big((\mu^*+\beta^*)\mathbf{H}\big)$ is equal to one since $\mu^*+\beta^*>0$. On the other hand,
$\mathbf{W}^*\ne\mathbf{0}$ is required to satisfy the minimum SINR requirement of the desired receiver  in C1 when $\Gamma_{\mathrm{req}}>0$. Therefore, $\Rank((\mu^*+\beta^*)\mathbf{H})=1$ and $\Rank(\mathbf{W}^*)=1$ when $\frac{\mathbf{D}^*_{\mathrm{C}_{2_k}}}{\Gamma_{\mathrm{tol}_k}}-\mathbf{D}^*_{\mathrm{C}_{5_k}}\succeq \mathbf{0},\forall k$ holds. \qed

\end{document}